\RequirePackage{lineno}
\documentclass[twocolumn,nofootinbib,floatfix,superscriptaddress]{revtex4-1}

\usepackage{amsmath,amssymb}
\usepackage{cancel}
\usepackage{graphicx}
\usepackage{epsfig}
\usepackage{amsmath,amsfonts,amssymb}
\usepackage{multirow}
\usepackage{colortbl}
\usepackage{hhline}
\usepackage{booktabs}
\usepackage{bbm}
\usepackage{tabulary}
\usepackage{pifont}
\usepackage{bbold}
\usepackage{pifont}
\newcommand{\cmark}{\ding{51}}
\newcommand{\xmark}{\ding{55}}
\newcommand\Tstrut{\rule{0pt}{0.6cm}}         
\newcommand\Bstrut{\rule[-3ex]{0pt}{0pt}}   
\newcommand\Tstrutt{\rule{0pt}{0.53cm}}        
\newcommand\Bstrutt{\rule[-2.5ex]{0pt}{0pt}}   
\newcommand\Tstruttt{\rule{0pt}{0.4cm}}        
\newcommand\Bstruttt{\rule[-1.2ex]{0pt}{0pt}}  
\newcolumntype{K}[1]{>{\centering\arraybackslash}m{#1}}
\newcommand{\specialcell}[2][c]{%
	\begin{tabular}[#1]{@{}c@{}}#2\end{tabular}}
\newcommand{\Mr}{\mathbf{M}_R}
\newcommand{\Yl}{\mathbf{Y}^\ell}
\newcommand{\Ynu}{\mathbf{Y}^\nu}
\newcommand{\Mnu}{\mathbf{M}^\nu}
\newcommand{\Ur}{\mathbf{U}_R}
\newcommand{\Unu}{\mathbf{U}_\nu}
\newcommand{\Ul}{\mathbf{U}_\ell}
\newcommand{\U}{\mathbf{U}}
\newcommand{\dmatm}{\Delta m^2_{31}}
\newcommand{\dmsol}{\Delta m^2_{21}}

\begin{document}

\title{The minimal type-I seesaw model with maximally-restricted texture zeros}

\author{D. M. Barreiros}
\affiliation{Departamento de F\'{\i}sica and CFTP, Instituto Superior T\'ecnico, Universidade de Lisboa, Lisboa, Portugal}
\author{R. G. Felipe}
\affiliation{Departamento de F\'{\i}sica and CFTP, Instituto Superior T\'ecnico, Universidade de Lisboa, Lisboa, Portugal}
 \affiliation{Instituto Superior de Engenharia de Lisboa, Rua Conselheiro Em\'{\i}dio Navarro, 1959-007 Lisboa, Portugal}
\author{F. R. Joaquim}
\affiliation{Departamento de F\'{\i}sica and CFTP, Instituto Superior T\'ecnico, Universidade de Lisboa, Lisboa, Portugal}


\begin{abstract}
In the context of Standard Model (SM) extensions, the seesaw mechanism provides the most natural explanation for the smallness of neutrino masses. In this work we consider the most economical type-I seesaw realization in which two right-handed neutrinos are added to the SM field content. For the sake of predictability, we impose the maximum number of texture zeros in the lepton Yukawa and mass matrices. All possible patterns are analyzed in the light of the most recent neutrino oscillation data, and predictions for leptonic CP violation are presented. We conclude that, in the charged-lepton mass basis, eight different texture combinations are compatible with neutrino data at $1\sigma$, all of them for an inverted-hierarchical neutrino mass spectrum. Four of these cases predict a CP-violating Dirac phase close to $3\pi/2$, which is around the current best-fit value from global analysis of neutrino oscillation data. If one further reduces the number of free parameters by considering three equal elements in the Dirac neutrino Yukawa coupling matrix, several texture combinations are still compatible with data but only at $3\sigma$. For all viable textures, the baryon asymmetry of the Universe is computed in the context of thermal leptogenesis, assuming (mildly) hierarchical heavy Majorana neutrino masses $M_{1,2}$. It is shown that the flavored regime is ruled out, while the unflavored one requires $M_{1} \sim 10^{14}$~GeV.

\end{abstract}

\maketitle

\section{Introduction}
\label{Intro}

The discovery of neutrino oscillations provided a solid evidence for physics beyond the Standard Model (SM), by confirming the existence of neutrino masses and mixing. From the theory viewpoint, the most straightforward and elegant way of accounting for them consists of adding right-handed (RH) neutrinos to the SM field content. If heavy enough, these states can mediate neutrino masses at the classical level through the well-known seesaw mechanism~\cite{Minkowski:1977sc}. Besides supplying an explanation for small neutrino masses, the addition of RH neutrinos to the SM allows for the leptogenesis mechanism~\cite{Fukugita:1986hr} to work through the out-of-equilibrium decays of the heavy neutrinos in the early Universe (for reviews see e.g.~\cite{Buchmuller:2004nz,Davidson:2008bu,Chun:2017spz,Fong:2013wr}). This offers an answer for another SM puzzle: the baryon asymmetry of the Universe (BAU). 

Although, in principle, the number of RH neutrinos is arbitrary, at least two are necessary to explain the present neutrino oscillation data, namely, three nonzero neutrino mixing angles and two mass-squared differences. Interestingly, at least two RH neutrinos are also required for leptogenesis to be realized. Therefore, the two RH neutrino seesaw model (2RHNSM) is not only a minimal model for neutrino masses, but also for the generation of the BAU in the context of leptogenesis. Still, even in this scenario, the number of parameters describing the neutrino Lagrangian at high energies is larger than the number of low-energy observables currently (or potentially) measured by experiments. One way of increasing predictability is to consider texture zeros in the lepton Yukawa and mass matrices, which can be motivated, for instance, by imposing U(1) Abelian flavor symmetries~\cite{Grimus:2004hf,Cebola:2015dwa}. In general, texture zeros imply predictions not only for low-energy neutrino parameters but also for the BAU, since leptogenesis is sensitive to the couplings which control neutrino masses and mixing. Therefore, a complete study of all possible texture zeros in the light of most recent neutrino data is welcome. In particular, since neutrino experiments are starting to deliver some information regarding leptonic CP violation~\cite{Branco:2011zb}, predictions for low-energy CP phases are of utmost importance. At the same time, a connection with leptogenesis can also be established in this framework~\cite{Branco:2011zb,Hagedorn:2017wjy}. These questions have already been partially covered in the literature. For instance, the compatibility of texture-zero hypothesis in the 2RHNSM with neutrino data has been studied in Refs.~\cite{Frampton:2002qc,Ibarra:2003up,Harigaya:2012bw,Rink:2016vvl,Shimizu:2017fgu} and, in the context of leptogenesis, in Refs.~\cite{GonzalezFelipe:2003fi,Joaquim:2005zv,Abada:2006ea,Zhang:2015tea,Siyeon:2016wro,Geib:2017bsw,Achelashvili:2017nqp,Shimizu:2017vwi}. 

In this work, we revisit the 2RHNSM in maximally restricted texture-zero scenarios, i.e. when the maximum number of texture zeros is imposed in the lepton Yukawa and mass matrices. Moreover, we consider cases in which equality relations among the Dirac neutrino Yukawa couplings exist. For textures that reproduce the observed neutrino mass and mixing patterns, we present the predictions for low-energy CP violation, neutrinoless double beta decay and the BAU. Special attention will be paid to the treatment of leptogenesis in the 2RHNSM. Contrary to what is usually done in the literature, where only the decay of the lightest heavy neutrino is considered, we include decays of both heavy neutrinos in our analysis. Moreover, flavor effects which arise from the fact that lepton interactions become out of equilibrium at different temperatures are taken into account.

This paper is organized as follows. In Section~\ref{sec1} we set the basics of the 2RHNSM, by describing the model and identifying the number of parameters at high and low energies. Afterwards, in Section~\ref{sec2}, the maximally-restricted texture zero matrices are identified, and their compatibility with neutrino data is analyzed. Furthermore, the predictions for Dirac and Majorana CP phases are shown, together with those for the effective neutrino mass parameter relevant for neutrinoless double beta decays. We also consider cases with three equal elements in the Dirac neutrino Yukawa coupling matrix in Section~\ref{sec3a}. We then compute the BAU in the thermal leptogenesis framework in Section~\ref{sec3}, and determine under which conditions its value is compatible with the observed one. Our conclusions are drawn in Section~\ref{sec4}.

\section{The two right-handed neutrino seesaw model}
\label{sec1}

Considering only Yukawa and mass terms, the lepton Lagrangian density for the SM extended with RH neutrino fields $\nu_{R}$ is $\mathcal{L}=\mathcal{L}_{\ell}+\mathcal{L}_\nu$ with
\begin{align}
	\mathcal{L}_\nu&=-\overline{\ell_{L}}\Ynu \tilde{\Phi}\nu_{R} -\frac{1}{2}\overline{(\nu_{R})^c}\Mr\nu_{R} + \text{H.c.}\,,	\label{LtypeI}\\
	\mathcal{L}_\ell&=-\overline{\ell_{L}}\Yl \Phi\,e_{R}  + \text{H.c.}\,.\label{Lcl}
\end{align}
Here, $\ell_{L}$ and $\Phi$ are the SM lepton and Higgs doublets, respectively, $\tilde{\Phi}=i\sigma_2 \Phi^\ast$, and $e_R$ denote the RH charged-lepton fields. The Dirac neutrino Yukawa couplings and RH neutrino mass matrices are described by $\Ynu$ and $\Mr$. For $N$ RH neutrinos, $\Ynu$ and $\Mr$ are $3\times N$ and $N\times N$ general complex matrices, being $\Mr$ symmetric. After integrating out the $\nu_R$'s, the effective Majorana neutrino mass matrix $\Mnu$, obtained upon electroweak symmetry breaking, is given by the seesaw formula~\cite{Minkowski:1977sc}
\begin{align}
\Mnu=-v^2\Ynu\Mr^{-1}{\Ynu}^T\,,
\label{Mnuseesaw}
\end{align}
which is valid for $\Mr\gg v$, where $v=174\,{\rm GeV}$ is the vacuum expectation value of the neutral component of $\Phi$. This (symmetric) matrix is diagonalized by a unitary matrix $\mathbf{U_\nu}$ as
\begin{align}
\Unu^T \Mnu \Unu=\text{diag}(m_1,m_2,m_3)\equiv \mathbf{d}_m\,,
\label{Mnudiag}
\end{align}
where $m_i$ are the (real and positive) effective neutrino masses. Considering that $\Ul$ rotates the left-handed (LH) charged-lepton fields to their diagonal mass basis, lepton mixing in charged currents is encoded in the so-called Pontecorvo-Maki-Nakagawa-Sakata (PMNS) unitary matrix $\U$ given by
\begin{align}
\U=\Ul^\dag \Unu\,.
\label{UPMNS}
\end{align}

Throughout this work we will use the standard parametrization~\cite{Patrignani:2016xqp}
\begin{widetext}
\begin{gather}
\mathbf{U}=\begin{pmatrix}
c_{12}c_{13}&s_{12}c_{13}&s_{13}e^{-i\delta}\\
-s_{12}c_{23}-c_{12}s_{23}s_{13}e^{i\delta}&c_{12}c_{23}-s_{12}s_{23}s_{13}e^{i\delta}&s_{23}c_{13}\\
s_{12}s_{23}-c_{12}c_{23}s_{13}e^{i\delta}&-c_{12}s_{23}-s_{12}c_{23}s_{13}e^{i\delta}&c_{23}c_{13}
\end{pmatrix}\!\!\begin{pmatrix}
1&0&0\\
0&e^{i\alpha_{21}/2}&0\\
0&0&e^{i\alpha_{31}/2}\\
\end{pmatrix}\,,
\label{Uparam}
\end{gather}
\end{widetext}
where $c_{ij}\equiv \cos\theta_{ij}$, $s_{ij}\equiv \sin\theta_{ij}$ and $\theta_{ij}\,(i < j=1,2,3)$ are the three lepton mixing angles. The phases $\delta$ and $\alpha_{21,31}$ are Dirac and Majorana-type CP-violating phases, respectively.

The present values for $\theta_{ij}$, $\delta$ and $\Delta m^2_{ij}=m_i^2-m_j^2$, extracted from global analyses of all neutrino oscillation data~\cite{deSalas:2017kay,Esteban:2016qun,Capozzi:2016rtj}, are given in Table~\ref{datatable} for both normally-ordered (NO) and inverted-ordered (IO) neutrino mass spectra defined as:
\begin{eqnarray}
{\text {NO}:}& m_1 < m_2 < m_3\;\, (\dmatm >0)\,,\\
{\text {IO}:}& m_3 < m_1 < m_2\;\, (\dmatm <0)\,.
\label{NOIO}
\end{eqnarray}
Notice that although neutrino mixing angles and mass-squared differences are known with very good precision, the experimental sensitivity to the value of $\delta$ is still limited, and the statistical significance of the presented ranges for that parameter is low.

\begin{table}[t!]
	\centering
	\setlength\extrarowheight{3pt}
	\begin{tabular}{ccc}
		\hline\hline
		\textbf{Parameter}&Best Fit $\pm1\sigma$&$3\sigma$ range\\
		\hline
		$\theta_{12}\;(^{\circ})$&$34.5_{-1.0}^{+1.1}$&$31.5\rightarrow38.0$\\[0.15cm]
		$\theta_{23}\;(^{\circ})$ [NO] &$41.0\pm1.1$&$38.3\rightarrow52.8$\\
		$\theta_{23}\;(^{\circ})$ [IO] &$50.5\pm1.0$&$38.5\rightarrow53.0$\\[0.15cm]
		$\theta_{13}\;(^{\circ})$ [NO] &$8.44_{-0.15}^{+0.18}$&$7.9\rightarrow8.9$\\
		$\theta_{13}\;(^{\circ})$ [IO] &$8.41_{-0.17}^{+0.16}$&$7.9\rightarrow8.9$\\[0.15cm]
		$\delta\;(^{\circ})$ [NO] &$252_{-36}^{+56}$&$0\rightarrow360$\\
		\multirow{2}{*}{$\delta\;(^{\circ})$ [IO]} &\multirow{2}{*}{$259_{+47}^{+41}$}&$0\rightarrow31$ \\
		 &&$142\rightarrow360$\\[0.15cm]
		$\Delta m_{21}^2\;(\times 10^{-5}\;\text{eV}^2)$&$7.56\pm0.19$&$7.05\rightarrow8.14$\\[0.15cm]
		$|\Delta m_{31}^2|\;(\times 10^{-3}\;\text{eV}^2)$ [NO] &$2.55\pm0.04$&$2.43\rightarrow2.67$\\
		$|\Delta m_{31}^2|\;(\times 10^{-3}\;\text{eV}^2)$ [IO] &$2.49\pm0.04$&$2.37\rightarrow2.61$\\
		\hline\hline
	\end{tabular}
\caption{Neutrino oscillation parameters obtained from the global analysis of Ref.~\cite{deSalas:2017kay} (see also Refs.~\cite{Esteban:2016qun} and \cite{Capozzi:2016rtj}).}
	\label{datatable}
\end{table} 

Let us now consider the simplest type-I seesaw model which can account for the data presented in Table~\ref{datatable}, i.e. the 2RHNSM. In this case, $\Ynu$ and $\Mr$ are $3\times 2$ and $2 \times 2$ matrices, respectively. In the mass-eigenstate basis of $\nu_R$, the free parameters in the Lagrangian (\ref{LtypeI}) are the two RH neutrino masses $M_{1,2}$, and the twelve real parameters of $\Ynu$. By rotating the LH charged-lepton fields, one is able to eliminate three parameters from $\Ynu$, leaving a total of eleven. Since for the 2RHNSM the effective neutrino mass matrix $\Mnu$ given in Eq.~(\ref{Mnuseesaw}) is rank two, $m_1=0\,(m_3=0)$ for NO (IO).\footnote{From now on we will denote these two cases by normal (NH) and inverted hierarchy (IH), respectively.} Moreover, the diagonal phase matrix in Eq.~(\ref{Uparam}) must be replaced by $\text{diag}(1,e^{i\alpha/2},1)$ since, in the presence of a massless neutrino, only one Majorana phase is physical. Thus, in the 2RHNSM, the low-energy neutrino sector is described by seven parameters (two masses, three mixing angles and two CP-violating phases), to be compared with the eleven at high energies. 

One convenient way of parameterizing $\Ynu$ relies on the so-called Casas-Ibarra parametrization~\cite{Casas:2001sr}. In the basis where both $\Mr$ and $\Yl$ are diagonal,
\begin{gather}
\mathbf{Y}^\nu=v^{-1}\U^*\,\mathbf{d}_m^{1/2}\,\mathbf{R}\,\mathbf{d}_M^{1/2}\,,
\label{CasasandIbarra}
\end{gather}
with $\mathbf{d}_M=\text{diag}(M_1,M_2)$. The matrix $\mathbf{R}$ is a $3\times 2$ complex orthogonal matrix which can be parametrized by a single complex angle $z$ in the following way
\begin{gather}
\mathbf{R}_{\text{NH}}=\begin{pmatrix}
0&0\\
\cos z&-\sin z\\
\xi \sin z&\xi \cos z 
\end{pmatrix}
\;,\;
\mathbf{R}_{\text{IH}}=\begin{pmatrix}
\cos z &-\sin z \\
\xi \sin z &\xi \cos z\\
0&0
\end{pmatrix}\; ,
\label{RmatrixIO}
\end{gather}
with $\xi=\pm 1$. Notice that, in the case of a non-diagonal $\Mr$, the right-hand side of Eq.~(\ref{CasasandIbarra}) must be multiplied on the right by $\Ur^\dag$, being $\Ur$ the unitary matrix which diagonalizes $\Mr$ as $\Ur^T \Mr \Ur=\mathbf{d}_M$. 

Clearly, even in the simplest minimal type-I seesaw model, there are more free independent parameters at high energies than at low energies. In order to reduce the degree of arbitrariness of the 2RHNSM, in the next section we will introduce maximally-restricted texture zeros and study their phenomenological implications.

\section{Maximally-restricted texture zeros}
\label{sec2}

In this section we will study the implications of imposing texture zeros in $\Yl$, $\Ynu$ and $\Mr$. Our guiding principle is to consider the maximum number of zeros such that the charged-lepton masses and neutrino data can be accommodated. In the former case, this corresponds to having six zeros in $\Yl$, which guarantees three non-degenerate masses. There are six textures of this type related among each other by permutations of rows and/or columns applied to $\Yl_{\rm diag}={\rm diag}(y_e,y_\mu,y_\tau)$, where $y_{e,\mu,\tau}=m_{e,\mu,\tau}/v$. Textures for $\Ynu$ with three or more zeros lead to vanishing mixing angles and/or two massless neutrinos, being therefore excluded experimentally. In principle, with two texture zeros in $\Ynu$, all neutrino data could be reproduced. There are fifteen different types of $3\times 2$ matrices with two vanishing entries. Some of them are automatically excluded by present neutrino data, namely,

\begin{itemize}
\item {Textures with two zeros placed in the same line $j$ of $\Ynu$ are excluded since these lead to the case in which the two RH neutrino fields are decoupled from the lepton flavor $j$. Therefore, all elements in line (and column) $j$ of the Majorana neutrino mass matrix $\Mnu$ vanish, implying the existence of two vanishing mixing angles $\theta_{ij}$, which is excluded by the data. In practice, this corresponds to the situation in which one neutrino flavor state coincides with its mass eigenstate.}

\item {If both zeros are placed in lines $(i,j)$ of the same column in $\Ynu$, then lines (and columns) $(i,j)$ of $\Mnu$ are linearly dependent. Thus, at least one mixing angle $\theta_{ij}$ is zero, leading to the unrealistic case in which one flavor eigenstate is a superposition of only two of the three mass eigenstates.}	
\end{itemize} 
We therefore conclude that the maximally-allowed number of texture zeros in $\Ynu$ is two. The $\Ynu$ textures to be analyzed are of the type:
\begin{gather}
\begin{matrix}
{\rm T}_1:\;\begin{pmatrix}
0&\times\\
\times&0\\
\times&\times
\end{pmatrix},&
{\rm T_2}: \;\begin{pmatrix}
0&\times\\
\times&\times\\
\times&0
\end{pmatrix},&
{\rm T_3}:\; \begin{pmatrix}
\times&\times\\
0&\times\\
\times&0
\end{pmatrix},\\
&&&\\
{\rm T_4}:\; \begin{pmatrix}
\times&0\\
0&\times\\
\times&\times
\end{pmatrix},&
{\rm T_5}: \;\begin{pmatrix}
\times&0\\
\times&\times\\
0&\times
\end{pmatrix},&
{\rm T_6}:\; \begin{pmatrix}
\times&\times\\
\times&0\\
0&\times
\end{pmatrix},
\end{matrix}
\label{Tstructures}
\end{gather}
where the symbol $\times$ denotes a generic non-vanishing entry.
\begin{table}[t!]
	\centering
	\setlength\extrarowheight{3pt}
	\begin{tabular}{K{1.5cm}|K{1.5cm}|c|K{1.2cm}|K{1.2cm}}
		\hline\hline
		$\Ynu$&$\Mr$&$\Mnu$&\textbf{NH}&\textbf{IH}\\
		\hline
		\multirow{1}{*}[-0.15cm]{T$_1$, T$_2$}&\multirow{1}{*}[-0.15cm]{R$_2$}&\multirow{2}{*}{A: $\begin{pmatrix}
			0 &\times&\times\\
			\cdot&\times&\times\\
			\cdot&\cdot&\times
			\end{pmatrix}$}&\multirow{2}{*}[-0.15cm]{\xmark}&\multirow{2}{*}[-0.25cm]{\xmark}\\[0.25cm]
		\cline{1-2}
		\multirow{1}{*}[-0.15cm]{T$_4$, T$_5$}&\multirow{1}{*}[-0.15cm]{R$_3$}&&\\[0.25cm]
		\hline
		T$_1$, T$_4$&R$_1$&B: $\begin{pmatrix}
		\times &0&\times\\
		\cdot&\times&\times\\
		\cdot&\cdot&\times
		\end{pmatrix}$&\xmark&\cmark (${1\sigma}$)\\
		\hline
		T$_2$, T$_5$&R$_1$&C: $\begin{pmatrix}
		\times &\times&0\\
		\cdot&\times&\times\\
		\cdot&\cdot&\times
		\end{pmatrix}$&\xmark&\cmark (${1\sigma}$)\\
		\hline
		\multirow{1}{*}[-0.15cm]{T$_3$, T$_4$}&\multirow{1}{*}[-0.15cm]{R$_2$}&\multirow{2}{*}{D: $\begin{pmatrix}
			\times &\times&\times\\
			\cdot&0&\times\\
			\cdot&\cdot&\times
			\end{pmatrix}$}&\multirow{2}{*}[-0.15cm]{\xmark}&\multirow{2}{*}[-0.25cm]{\cmark (${1\sigma}$)}\\[0.25cm]
		\cline{1-2}
		\multirow{1}{*}[-0.15cm]{T$_1$, T$_6$}&\multirow{1}{*}[-0.15cm]{R$_3$}&&\\[0.25cm]
		\hline
		T$_3$, T$_6$&R$_1$&E: $\begin{pmatrix}
		\times &\times&\times\\
		\cdot&\times&0\\
		\cdot&\cdot&\times
		\end{pmatrix}$&\xmark&\xmark\\
		\hline
		\multirow{1}{*}[-0.15cm]{T$_5$, T$_6$}&\multirow{1}{*}[-0.15cm]{R$_2$}&\multirow{2}{*}{F: $\begin{pmatrix}
			\times &\times&\times\\
			\cdot&\times&\times\\
			\cdot&\cdot&0
			\end{pmatrix}$}&\multirow{2}{*}[-0.15cm]{\xmark}&\multirow{2}{*}[-0.25cm]{\cmark (${3\sigma}$)}\\[0.25cm]
		\cline{1-2}
		\multirow{1}{*}[-0.15cm]{T$_2$, T$_3$}&\multirow{1}{*}[-0.15cm]{R$_3$}&&\\[0.25cm]
		\hline\hline
	\end{tabular}
	\caption{Textures for the effective neutrino mass matrix $\Mnu$ (third column) obtained with the seesaw formula given in Eq.~(\ref{Mnuseesaw}), and considering the textures T$_1$-T$_6$ for $\Ynu$ (first column) and R$_1$-R$_3$ for $\Mr$ (second column). The check (\cmark) and cross (\xmark) marks indicate whether the texture combination is compatible or not with data.}
	\label{tabR1R2R3}
\end{table}

As for $\Mr$, with more that two texture zeros, at least one of the RH neutrinos is massless. On the other hand, with one texture zero, there are three different patterns for $\Mr$:
\begin{align}
{\rm R}_1:\begin{pmatrix}
\times&0\\
\cdot&\times
\end{pmatrix}\;,\;
{\rm R_2}:\begin{pmatrix}
0&\times\\
\cdot&\times
\end{pmatrix}\;,\;
{\rm R_3}:\begin{pmatrix}
\times&\times\\
\cdot&0
\end{pmatrix}\,,
\label{Rstructures}
\end{align}
with the dot $(\cdot)$ indicating the symmetric nature of the matrix.
Combining them with the $\Ynu$ textures (\ref{Tstructures}) through the seesaw formula (\ref{Mnuseesaw}), one obtains the textures for $\Mnu$ given in the third column of Table~\ref{tabR1R2R3}. All cases A-F feature the presence of one texture zero in $\Mnu$. Notice that sets of $(\Ynu,\Mr)$ textures related by simultaneous permutations of the columns in $\Ynu$, and lines and columns in $\Mr$, lead to the same $\Mnu$ due to invariance of Eq.~(\ref{Mnuseesaw}) under $\nu_R$ rotations. Moreover, when $\Mr$ is diagonal (texture R$_1$), $\Mnu$ is the same for $\Ynu$ textures related by a column permutation. For instance, the sets $(\mathrm{T}_1,\mathrm{R}_1)$ and $(\mathrm{T}_4,\mathrm{R}_1)$ lead to the same low-energy predictions since $\mathrm{T}_1$ and $\mathrm{T}_4$ are related by column permutation.

The condition $\Mnu_{\alpha\beta}=0$ imposes relations among the neutrino parameters. In particular, from Eq.~(\ref{Mnudiag}) it is straightforward to conclude that~\cite{Xing:2003ic,Gautam:2015kya}
\begin{eqnarray}
\text{NH}:&\dfrac{m_2}{m_3}&\!\!=-\dfrac{\U^*_{\alpha 3}\U^*_{\beta 3}}{\U^*_{\alpha 2}\U^*_{\beta 2}}\;,
\label{NHrelation}\\
\text{IH}:&\dfrac{m_1}{m_2}&\!\!=-\dfrac{\U^*_{\alpha 2}\U^*_{\beta 2}}{\U^*_{\alpha 1}\U^*_{\beta 1}}\;.
\label{IHrelation}
\end{eqnarray}
%
Taking into account that neutrino masses $m_i$ are real and positive, $m_2^2=\dmsol$ ($m_2^2=\dmsol+|\dmatm|$) and $m_3^2=\dmatm$ ($m_1^2=|\dmatm|$) for NH (IH). Thus, we have
\begin{eqnarray}
\text{NH}:\;\;\;\;\;\;\;r_\nu&=&\left|\dfrac{\U^*_{\alpha 3}\U^*_{\beta 3}}{\U^*_{\alpha 2}\U^*_{\beta 2}}\right|^2,
\label{NHrelationb}\\
\text{IH}:\dfrac{1}{1+r_\nu}&=&\left|\dfrac{\U^*_{\alpha 2}\U^*_{\beta 2}}{\U^*_{\alpha 1}\U^*_{\beta 1}}\right|^2,\;\;\;r_\nu\equiv \dfrac{\dmsol}{|\dmatm|}\,.
\label{IHrelationb}
\end{eqnarray}

Given the parametrization in Eq.~(\ref{Uparam}), and the experimentally-allowed ranges for the mixing angles presented in Table~\ref{datatable}, one can test which textures lead to viable values of $r_\nu$ using the above relations. From all cases, the simplest one to be analyzed is texture A, for which $r_\nu$ is simply given by
\begin{align}
{\rm NH:} \,\,r_\nu=\frac{t_{13}^4}{s_{12}^4}\simeq 0.005\;,\; 
{\rm IH:}\,\, r_\nu=\frac{1}{t_{12}^4}-1\simeq 3.5\,.
\end{align}
These numerical estimates, obtained using the best-fit values given in Table~\ref{datatable}, indicate that texture A is disfavored by data, independently of the value of $\delta$.

\begin{figure*}[t!]
	\includegraphics[scale=0.65]{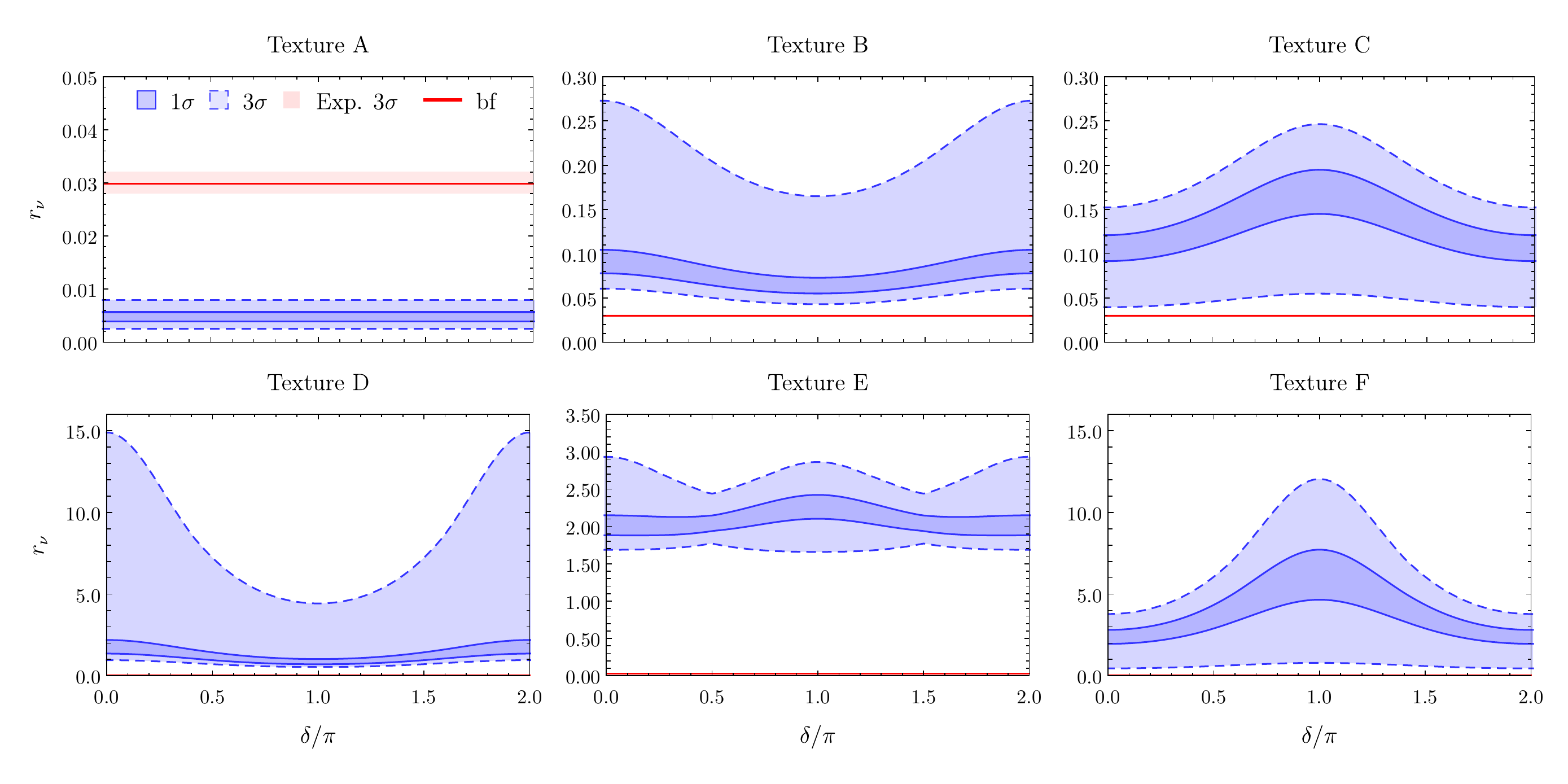}
	\caption{Predictions for $r_\nu$ as a function of $\delta$ in the NH case, using the $3\sigma$ (light blue) and $1\sigma$ (dark blue) ranges given in Table~\ref{datatable} for the mixing angles $\theta_{ij}$. The horizontal pink band (red line) denotes the $3\sigma$ range (best-fit value) for $r_\nu$ (see Eq.~\eqref{IHrelationb}), obtained using the data of Table~\ref{datatable}.}
	\label{NHmassratios}
\end{figure*}

\begin{figure*}[t!]
	\includegraphics[scale=0.65]{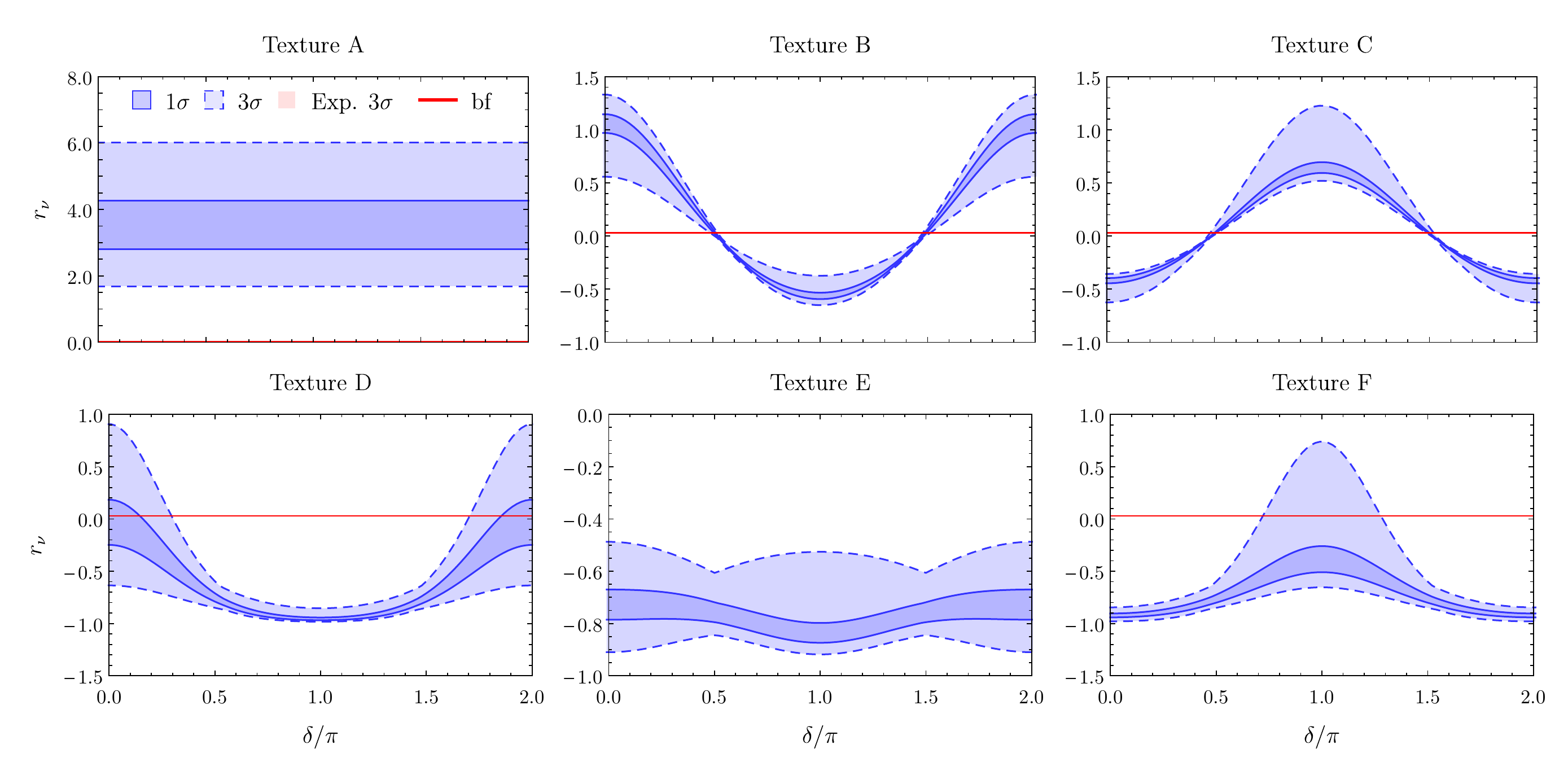}
	\caption{Predictions for $r_\nu$ as a function of $\delta$ in the IH case, using the $3\sigma$ (light blue) and $1\sigma$ (dark blue) ranges given in Table~\ref{datatable} for the mixing angles $\theta_{ij}$. The horizontal pink band (red line) denotes the $3\sigma$ range (best-fit value) for $r_\nu$ (see Eq.~\eqref{IHrelationb}), obtained using the data of Table~\ref{datatable}.}
	\label{IHmassratios}
\end{figure*}

By varying the mixing angles in their experimentally $1\sigma$ and $3\sigma$ allowed regions,\footnote{We will perform our analysis considering a diagonal charged-lepton Yukawa matrix $\Yl_{\rm diag}$. In the end of this section, we will comment on how the results change when  the remaining five $\Yl$ textures with six zeros are considered.} we plot $r_\nu$ as a function of $\delta$ in Figs.~\ref{NHmassratios} and \ref{IHmassratios} for NH  and IH, respectively, using Eqs.~(\ref{NHrelationb}) and (\ref{IHrelationb}) together with Eq.~(\ref{Uparam}). In light (dark) blue we show the $r_\nu$ regions obtained when all mixing angles vary in their $3\sigma$ ($1\sigma$) experimental ranges. The horizontal pink bands (red line) indicate the $3\sigma$ experimental range (best-fit value) for $r_\nu$. From these results, we conclude that all textures with one zero in $\Mnu$ are incompatible with neutrino data at more than $3\sigma$ level for NH. In the context of the 2RHNSM with texture zeros in $\Ynu$ and $\Mr$, this means that all combinations shown in Table~\ref{tabR1R2R3} are excluded for that type of neutrino mass spectrum. For IH (Fig.~\ref{IHmassratios}) and specific ranges of $\delta$, one obtains values for $r_\nu$ compatible with the data at $1\sigma$ for textures B, C and D, and only at $3\sigma$ for texture F. Therefore, all combinations of textures for $\Ynu$ and $\Mr$ leading to textures B, C, D and F for $\Mnu$ are viable. Notice that only textures B and C predict $r_\nu$ values in its $1\sigma$ range, for $\delta$ around its best-fit value.
\begin{table}
	\centering
	\setlength\extrarowheight{3pt}
	\begin{tabular}{ll}
		\hline\hline
		$\Mnu\;$ &$\hspace{2cm}$\textbf{CP-violating phases}\\
		\hline 
		\multirow{3}{*}{B}& $c_\delta=2\dfrac{[s^4_{12}(1+r_\nu)-c^4_{12}]s^2_{23}s^2_{13}+r_\nu c^2_{23}s_{12}^2c^2_{12}}{[s^2_{12}(1+r_\nu)+c^2_{12}]\sin(2\theta_{12})\sin(2\theta_{23})s_{13}}$\Tstrut\Bstrut\\
		& $c_\alpha=\dfrac{(2+r_\nu)c^2_{23}s^2_{12}c_{12}^2-[s^4_{12}(1+r_\nu)+c^4_{12}]s^2_{23}s^2_{13}}{2\sqrt{1+r_\nu}(c^2_{23}+s^2_{23}s^2_{13})s^2_{12}c^2_{12}}$\Tstrut\Bstrut\\
		\hline
		\multirow{3}{*}{C}& $c_\delta=-2\dfrac{[s^4_{12}(1+r_\nu)-c^4_{12}]c^2_{23}s^2_{13}+r_\nu s^2_{23}s_{12}^2c^2_{12}}{[s^2_{12}(1+r_\nu)+c^2_{12}]\sin(2\theta_{12})\sin(2\theta_{23})s_{13}}$\Tstrut\Bstrut\\
		&$c_\alpha=\dfrac{(2+r_\nu)s^2_{23}s^2_{12}c_{12}^2-[s^4_{12}(1+r_\nu)+c^4_{12}]c^2_{23}s^2_{13}}{2\sqrt{1+r_\nu}(s^2_{23}+c^2_{23}s^2_{13})s^2_{12}c^2_{12}}$\Tstrut\Bstrut\\
		\hline
		\multirow{3}{*}{D}& $c_\delta=2\dfrac{(c^2_{12}\sqrt{1+r_\nu}-s^2_{12})c^2_{23}+(s^2_{12}\sqrt{1+r_\nu}-c^2_{12})s^2_{23}s^2_{13}}{(\sqrt{1+r_\nu}+1)\sin(2\theta_{12})\sin(2\theta_{23})s_{13}}$\Tstrut\Bstrut\\
		&$c_\alpha\simeq-\dfrac{3+\cos(4\theta_{12})-16s_{13}^2t_{23}^2}{2\sin^2(2\theta_{12})}$\Tstrut\Bstrut\\
		\hline
		\multirow{3}{*}{F}& $c_\delta=2\dfrac{(s^2_{12}-c^2_{12}\sqrt{1+r_\nu})s^2_{23}+(c^2_{12}-s^2_{12}\sqrt{1+r_\nu})c^2_{23}s^2_{13}}{(\sqrt{1+r_\nu}+1)\sin(2\theta_{12})\sin(2\theta_{23})s_{13}}$\Tstrut\Bstrut\\
		&$c_\alpha\simeq-\dfrac{3t_{23}^2+t_{23}^2\cos(4\theta_{12})+16s_{13}^2}{2t_{23}^2\sin^2(2\theta_{12})}$\Tstrut\Bstrut\\
		\hline\hline
	\end{tabular}
	\label{CPtable}
	\caption{Expressions for $\cos\delta\equiv c_\delta$ and $\cos\alpha \equiv c_\alpha$ for textures B, C, D and F. }
\end{table}

\begin{figure}
	\includegraphics[scale=0.65]{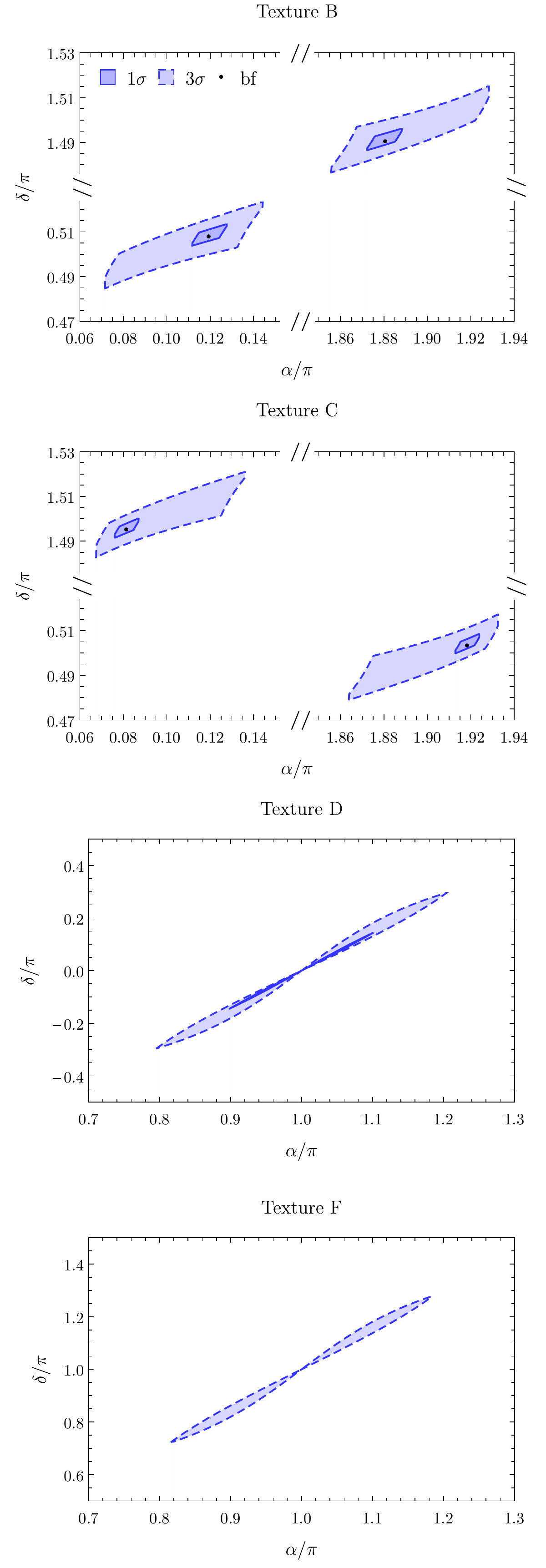} \\
	\caption{Predictions for the low-energy phases $\delta$ and $\alpha$ for textures B, C, D and F, using the $3\sigma$ (light blue) and $1\sigma$ (dark blue) ranges given in Table~\ref{datatable} for the mixing angles and neutrino mass-squared differences. The black dot corresponds to the predictions obtained with the best-fit values of $~\theta_{ij}$, $\dmsol$ and $|\dmatm|$.}
	\label{deltaalpha}
\end{figure}

Having identified the compatible textures, we now obtain expressions for $\delta$ in terms of the mixing angles and $r_\nu$ using Eq.~(\ref{IHrelationb}). By imposing that the right-hand side of Eq.~(\ref{IHrelation}) is real, we can obtain analytical expressions for the Majorana phase $\alpha$ as a function of $\theta_{ij}$, $r_\nu$ and $\delta$. In Table~\ref{CPtable}, we present the results for $c_\delta\equiv\cos\delta$ and $c_\alpha\equiv\cos\alpha$ for textures B, C, D and F when the neutrino mass spectrum is of IH type. It is worth mentioning that, although of different nature, $\delta$ and $\alpha$ are not independent phases in our case. This is due to the presence of zeros in the effective neutrino mass matrix.
Taking $\theta_{ij}$, $\dmsol$ and $\dmatm$ in their $3\sigma$ $(1\sigma)$ experimental ranges, we show in Fig.~\ref{deltaalpha} the light (dark) blue allowed regions in the $(\alpha,\delta)$ parameter space for textures B, C, D and F of $\Mnu$. We conclude that, for textures B and C, values of $\delta \simeq 3 \pi/2$ close to the best-fit value are allowed (cf. Table~\ref{datatable}). For such values of $\delta$, $\alpha \simeq 1.9 \pi\,(0.08\pi)$ is predicted for texture B (C). In fact, for these textures
\begin{align}
{\rm B:}\;&c_\delta \simeq \frac{r_\nu \sin(2\theta_{12})}{4s_{13}t_{23}}-\frac{s_{13}t_{23}}{\tan(2\theta_{12})}\,,\\
{\rm C:}\;&c_\delta \simeq -\frac{r_\nu t_{23} \sin(2\theta_{12})}{4s_{13}}+\frac{s_{13}}{t_{23}\tan(2\theta_{12})},
\end{align}
from which we see that $|c_\delta|\ll 1$, implying $\delta \simeq \pm \pi/2$. Instead, for textures D and F
\begin{align}
{\rm D:}\;&c_\delta \simeq \frac{1}{2s_{13}t_{23}\tan(2\theta_{12})}\,,\\
{\rm F:}\;&c_\delta \simeq -\frac{t_{23}}{2s_{13}\tan(2\theta_{12})}\,,
\end{align}
one obtains $|c_\delta| \sim \mathcal{O}(1)$ meaning that $\delta$ is far from $\pm \pi/2$. Therefore, as anticipated above, only textures B and C lead to $\delta$ values within the $1\sigma$ range of Table~\ref{datatable}. For textures D and F, the obtained values for $\delta$ are out of the $1\sigma$ range, but still within the $3\sigma$ one. 

\begin{figure}
	\includegraphics[scale=0.65]{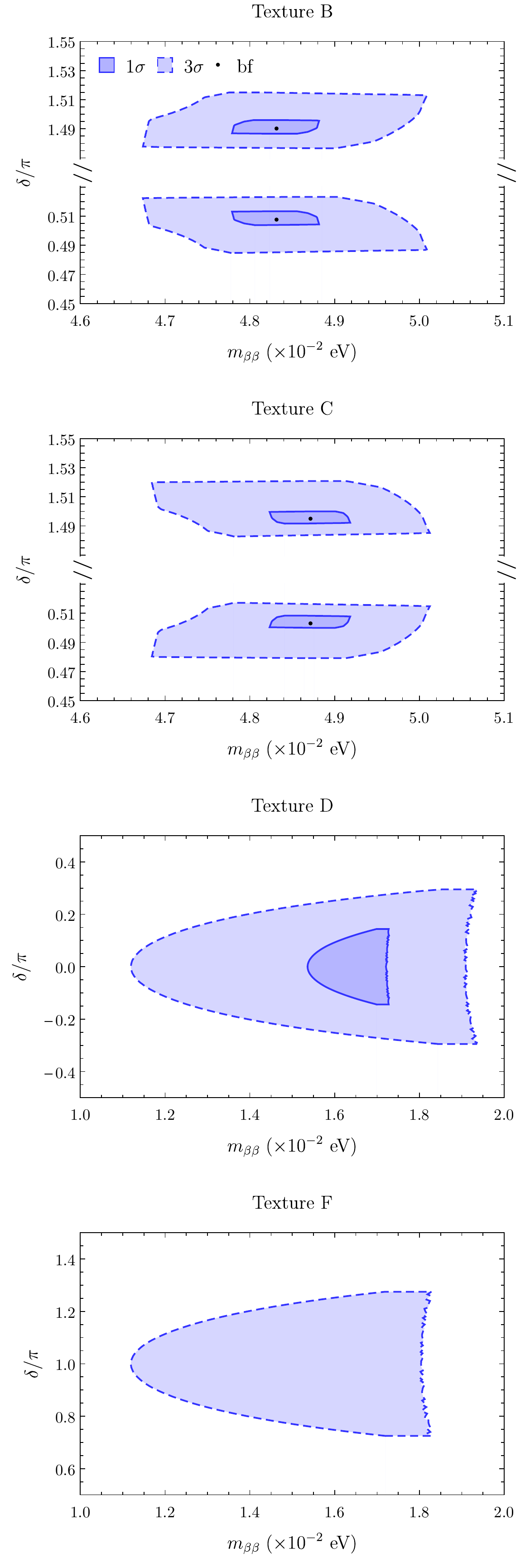} \\
	\caption{Predictions for $\delta$ and $m_{\beta\beta}$ for textures B, C, D and F, using the $3\sigma$ (light blue) and $1\sigma$ (dark blue) ranges given in Table~\ref{datatable} for the mixing angles and neutrino mass-squared differences. The black dot corresponds to the predictions obtained with the best-fit values of $~\theta_{ij}$, $\dmsol$ and $|\dmatm|$.}
	\label{mbbdelta}
\end{figure}

Presently, attempts to probe the Majorana nature of neutrinos are mainly based on neutrinoless double beta decay ($0\nu\beta\beta$) experiments. The observation of $0\nu\beta\beta$ decay would also provide a measurement of the neutrino mass scale, since the rate of this process is related to the square of the neutrino mass. A relevant quantity for $0\nu\beta\beta$ decay is the effective mass $m_{\beta\beta}$, which, for an IH neutrino mass spectrum, is given by
\begin{align}
m_{\beta\beta}&=\left|\sum_{i=1}^{3}m_i \U_{1i}^2\right|\nonumber\\
&=c_{13}^2|\dmatm|^{1/2}\,\left|c_{12}^2+(1+r_\nu)^{1/2}\,s_{12}^2e^{i\alpha} \right|.
\end{align}
 Given that $\alpha$ is a function of $\theta_{ij}$, $\delta$ and $r_\nu$, in Fig.~\ref{mbbdelta}, we show the allowed regions in the $(m_{\beta\beta},\delta)$-plane, taking into account the experimental ranges for the neutrino parameters (the color codes are the same used in  previous figures). The results are presented for textures B, C, D and F, where one can see that the value of $m_{\beta\beta}$ is around 50~meV (15~meV) for textures B and C (D and F). These values are compatible with all constraints coming from  $0\nu\beta\beta$ decay and cosmological experiments~\cite{Capozzi:2017ipn} for IH, but lie out of the sensitivity range of leading experiments like EXO-200~\cite{Albert:2017qto}, KamLAND-Zen~\cite{KamLAND-Zen:2016pfg}, GERDA~\cite{Agostini:2017dxu} and CUORE-0~\cite{Alduino:2016vtd}. Nevertheless, next-generation experiments will be able to test the IH spectrum (for a general discussion about future prospects and sensitivities of  $0\nu\beta\beta$ decay experiments see e.g. Ref.~\cite{Ostrovskiy:2016uyx}). 

In the above analysis, we have studied the cases with one texture zero in $\Mr$. Notice, however, that the maximally-allowed number of zeros in this matrix is actually two, leading to a single possible texture
\begin{align}
{\rm R_4:\,}\begin{pmatrix}
0&\times\\
\cdot&0
\end{pmatrix},
\label{R4text}
\end{align}   
which is characterized by a spectrum with two degenerate RH neutrinos. Combining through the seesaw formula (\ref{Mnuseesaw}) the matrix R$_4$ with all $\Ynu$ textures presented in Eq.~\eqref{Tstructures}, one obtains the textures for $\Mnu$ given in the third column of Table~\ref{tabR4}. One can see that in all cases $\Mnu$ contains two zeros, which have been tested individually above.\footnote{Analyses of $\Mnu$ with two texture zeros have been presented in Refs.~\cite{Cebola:2015dwa,Xing:2002ap,Dev:2006qe,Ludl:2011vv,Fritzsch:2011qv,Meloni:2012sx,Grimus:2012zm,Dev:2014dla,Kitabayashi:2015jdj} for the general case $m_{1,2,3}\neq 0$.} Moreover, additional relations among the elements of $\Mnu$ (see fourth column of Table~\ref{tabR4}) arise due to the specific form of $\Mr$, which contains a single parameter. For NH, all cases with ${\rm R}_4$ are excluded, since all textures with one zero in $\Mnu$  were already shown to be incompatible with data (see Table~\ref{tabR1R2R3}). For IH, combinations leading to textures ${\rm A}_1$ and ${\rm A}_2$ for $\Mnu$ are excluded due to the condition $\Mnu_{11}=0$ (see Table~\ref{tabR1R2R3}). 
\begin{table}
	\centering
	\setlength\extrarowheight{3pt}
	\begin{tabular}{K{1.2cm}|K{0.7cm}|c|K{2.1cm}|K{0.7cm}|K{0.7cm}}
		\hline\hline
		$\Ynu$&$\Mr$&$\Mnu$&\textbf{Relation in }$\Mnu$&\textbf{NH}&\textbf{IH}\\
		\hline
		T$_1$, T$_4$&\multirow{3}{*}[-1.5cm]{R$_4$}&A$_1$: $\begin{pmatrix}
		0 &\times&\times\\
		\cdot&0&\times\\
		\cdot&\cdot&\times
		\end{pmatrix}$&$\dfrac{\Mnu_{33}}{2\Mnu_{23}}=\dfrac{\Mnu_{13}}{\Mnu_{12}}$&\xmark&\xmark\\
		\cline{1-1}\cline{3-6}
		T$_2$, T$_5$&&A$_2$: $\begin{pmatrix}
		0 &\times&\times\\
		\cdot&\times&\times\\
		\cdot&\cdot&0
		\end{pmatrix}$&$\dfrac{\Mnu_{22}}{2\Mnu_{23}}=\dfrac{\Mnu_{12}}{\Mnu_{13}}$&\xmark&\xmark\\
		\cline{1-1}\cline{3-6}
		T$_3$, T$_6$&&D$_1$ : $\begin{pmatrix}
		\times &\times&\times\\
		\cdot&0&\times\\
		\cdot&\cdot&0
		\end{pmatrix}$&$\dfrac{\Mnu_{11}}{2\Mnu_{12}}=\dfrac{\Mnu_{13}}{\Mnu_{23}}$&\xmark&\xmark\\
		\hline\hline
	\end{tabular}
	\caption{Textures for the effective neutrino mass matrix $\Mnu$ (third column) obtained with the seesaw formula given in Eq.~(\ref{Mnuseesaw}), and considering the textures for $\Ynu$ (first column) and $\Mr$ (second column) presented in Eqs.~(\ref{Tstructures}) and (\ref{Rstructures}).}
	\label{tabR4}
\end{table}
As for texture D$_1$, although the conditions $\Mnu_{22}=0$ and $\Mnu_{33}=0$ are individually compatible with the data at $3\sigma$, they cannot be simultaneously verified, as one can see in Fig.~\ref{IHmassratios}, comparing the results for textures D and F. Indeed, from these plots one concludes that there is no overlap between the regions allowed by the data for the same values of $\delta$. This seems to contradict previous results obtained in the literature which state that textures with $\Mnu_{22}=\Mnu_{33}=0$ are compatible with the data (see e.g. Ref~\cite{Cebola:2015dwa}). Notice, however, that the results in those references were obtained for a general neutrino spectrum with $m_{1,2,3}\neq 0$. One can understand why texture D$_1$ in our case ($m_3=0$) is not valid by inspecting the relations between neutrino masses and $\U$ when the conditions $\Mnu_{22}=\Mnu_{33}=0$ are imposed, namely~\cite{Fritzsch:2011qv},
\begin{align}
\frac{m_3}{m_1}=\left|\frac{\U_{22}^2\U_{31}^2-\U_{21}^2\U_{32}^2}{\U_{23}^2\U_{32}^2-\U_{22}^2\U_{33}^2}\right|\,,\\
\frac{m_3}{m_2}=\left|\frac{\U_{22}^2\U_{31}^2-\U_{21}^2\U_{32}^2}{\U_{21}^2\U_{33}^2-\U_{23}^2\U_{31}^2}\right|\,.
\end{align}   
Therefore, if $m_3=0$ the condition 
\begin{align}
\left|\U_{22}^2\U_{31}^2-\U_{21}^2\U_{32}^2\right|=0
\end{align}
must be verified for texture D$_1$. The above relation can be approximately written as
\begin{align}
c_\delta\simeq \dfrac{2\cos(2\theta_{12})\cos(2\theta_{23})\pm\sqrt{2}\sqrt{\cos(4\theta_{12})+\cos(4\theta_{23})}}{4\sin(2\theta_{12})\sin(2\theta_{23})s_{13}}\,,
\end{align}
which, taking into account the current mixing angle data, always leads to a complex $c_\delta$. 

In conclusion, we have analyzed all possible textures with six zeros in $\Yl$, two zeros in $\Ynu$, and one or two zeros in $\Mr$. The compatibility of all textures is summarized in the last two columns of Tables~\ref{tabR1R2R3} and \ref{tabR4}, for NH and IH. We remark that no restriction has been imposed in the non-zero elements of those matrices. 

The results presented above are valid in the basis where $\Yl={\rm diag}(y_e,y_\mu,y_\tau)\equiv \Yl_{\rm diag}$ so that the charged lepton mass matrix is $\mathbf{M}^\ell={\rm diag}(m_e,m_\mu,m_\tau)$. One may wonder whether these conclusions hold for any other $\Yl$ with six zeros and only three nonzero elements. First, it is straightforward to see that any two nonzero elements in the same line/column lead to a massless charged lepton. This leaves us with six viable textures for $\Yl$ with six zeros, which can be obtained from $\Yl_{\rm diag}$ by applying permutations of lines and/or columns:
\begin{gather}
\begin{matrix}
{\rm L}_1:\;\begin{pmatrix}
\times&0 &0\\
0&\times &0\\
0&0 &\times
\end{pmatrix},&
{\rm L_2}: \;\begin{pmatrix}
0&\times &0\\
\times&0 &0\\
0&0 &\times
\end{pmatrix},&
{\rm L_3}:\; \begin{pmatrix}
0&0 &\times\\
0&\times &0\\
\times&0 &0
\end{pmatrix},\\
&&&\\
{\rm L}_4:\;\begin{pmatrix}
\times&0 &0\\
0&0 &\times\\
0&\times &0
\end{pmatrix},&
{\rm L_5}: \;\begin{pmatrix}
0&0 &\times\\
\times&0 &0\\
0&\times &0
\end{pmatrix},&
{\rm L_6}:\; \begin{pmatrix}
0&\times &0\\
0&0 &\times\\
\times&0 &0
\end{pmatrix}.
\end{matrix}
\label{Ylstructures}
\end{gather}
Obviously, if only column permutations (rotation of RH charged-lepton fields) are performed, then the results for a specific set of $\Ynu$ and $\Mr$ textures remain unchanged. However, if a permutation of the lines $i$ and $j$ in $\Yl$ is involved (rotation of LH charged-lepton fields by the permutation matrix $\mathbf{P}_{ij}$), then the same line permutation has to be performed in $\Ynu$. At the effective level, this corresponds to permuting the lines and columns $i$ and $j$ in the effective neutrino mass matrix $\Mnu$. Under these rotations, textures T$_1$-T$_6$ of $\Ynu$ and, consequently, A-F of $\Mnu$, are transformed among themselves. Thus, even if a given texture pair ($\Ynu$,$\Mr$) is not compatible with data in the $\Yl_{\rm diag}$ basis, this may not be the case in another $\Yl$ basis obtained from a line permutation $\mathbf{P}_{ij}$. 

To check the viability of a given set of textures $(\Yl,\Ynu,\Mr;\Mnu)=({\rm L}_i,{\rm T}_i,{\rm R}_i;\text{A-F})$ one has to identify the permutation $\mathbf{P}_{ij}$ which brings $\text{L}_i$ to $\Yl_{\rm diag}$, and find the transformed $\Mnu$ texture. For instance, consider the case $(\Yl_{\rm diag},{\rm T}_3,{\rm R}_1;{\rm E})$, shown in Table~\ref{tabR1R2R3} to be incompatible with data. Under $\mathbf{P}_{13}$, $\Yl_{\rm diag}$ is transformed into ${\rm L}_3$, while texture E becomes texture B, which is compatible with data at $1\sigma$. Therefore, although the set $(\Yl_{\rm diag},{\rm T}_3,{\rm R}_1;{\rm E})$ is not viable, the set $({\rm L}_3,{\rm T}_3,{\rm R}_1;{\rm E})$ is, since it corresponds to $(\Yl_{\rm diag},{\rm T}_4,{\rm R}_1;{\rm B})$ under $\mathbf{P}_{13}$. In Table~\ref{permut} we summarize the transformation properties of each $\Mnu$ texture under line permutations $\mathbf{P}_{ij}$, identifying in each case the compatibility with data taking into account the results obtained for $\Yl_{\rm diag}$ given in Table~\ref{tabR1R2R3}.

Notice that when $\Mr$ is of type $\text{R}_4$, the results presented in Table~\ref{tabR4} are valid for any $\Yl$ texture of type $\text{L}_i$. This is due to the fact that, under any permutation of lines and/or columns in $\Yl$, textures A$_{1,2}$ and D$_1$ (which are all excluded by data) transform among themselves.
\begin{table}[t]
	\setlength\extrarowheight{2pt}
	\begin{tabular}{K{1.0cm}l|K{1cm}l|K{1cm}l|K{1cm}l}
		\hline\hline
		\multicolumn{2}{c}{\textbf{Texture}}&\multicolumn{2}{|c}{$\mathbf{P}_{12}$}&\multicolumn{2}{|c}{$\mathbf{P}_{13}$}&\multicolumn{2}{|c}{$\mathbf{P}_{23}$}\\
		\hline
		A&\xmark&D&\cmark($1\sigma$)&F&\cmark($3\sigma$)&A&\xmark\\
		B&\cmark($1\sigma$)&B&\cmark($1\sigma$)&E&\xmark&C&\cmark($1\sigma$)\\
		C&\cmark($1\sigma$)&E&\xmark&C&\cmark($1\sigma$)&B&\cmark($1\sigma$)\\
		D&\cmark($1\sigma$)&A&\xmark&D&\cmark($1\sigma$)&F&\cmark($3\sigma$)\\
		E&\xmark&C&\cmark($1\sigma$)&B&\cmark($1\sigma$)&E&\xmark\\
		F&\cmark($3\sigma$)&F&\cmark($3\sigma$)&A&\xmark&D&\cmark($1\sigma$)\\		
		\hline\hline
	\end{tabular}
\caption{Transformation properties of $\Mnu$ (textures A-F) under $\mathbf{P}_{ij}$, which corresponds to permutations of the charged-lepton flavors $i$ and $j$. The compatibility of each texture with data is also indicated considering the results shown in Table~\ref{tabR1R2R3} for the case $\Yl=\Yl_{\rm diag}$.}
\label{permut}
\end{table}

\subsection{Imposing relations among the elements of $\Ynu$}
\label{sec3a}
 
We now intend to further restrict the two texture zero patterns analyzed above by imposing equality relations among the elements of $\Ynu$. The first obvious choice would be to consider all elements in $\Ynu$ to be equal. However, one can show that the eigenvector associated to $m_3=0$ is always $v_3=(\mp 1,-1,1)/\sqrt{3}$, leading to $s_{13}=\pm 1/\sqrt{3}$, which is excluded by the data. Thus, we move to the analysis of textures with two zeros in $\Ynu$ and three equal elements. Each case will be denoted by the labels of $\Ynu$, $\Mr$ and corresponding $\Mnu$ (see first column of Table~\ref{Tabrel}), and indexes of the $\Ynu$ equal elements (see second column of Table~\ref{Tabrel}). For instance, the cases with $\Ynu_{21}=\Ynu_{31}=\Ynu_{12}$ are denoted by $(21,31,12)$. Due to the highly constrained form of the involved matrices, extra relations among the elements of $\Mnu$ arise. These are shown in the third column of Table~\ref{Tabrel} for all possible combinations. Compatibility with neutrino data is determined by checking whether those relations are verified taking the allowed ranges for the neutrino parameters given in Table~\ref{datatable}. Also notice that the heavy Majorana neutrino masses and the elements of $\Mnu$ are related. In particular, defining the ratio
\begin{table*}[t]
	\centering
	\setlength\extrarowheight{3pt}
	\begin{tabular}{K{2.4cm}|K{1.6cm}|K{3.6cm}|K{3.8cm}|K{0.8cm}|c|c}
		\hline\hline
		($\Ynu$, $\Mr$, $\Mnu$)&\textbf{Equal elements in }$\Ynu$& \textbf{Relations in} $\Mnu$&$r_N\equiv M_2/M_1$&\textbf{IH}&\specialcell{\textbf{Low energy predictions}\\$(\theta_{12},\theta_{23},\theta_{13})^\circ$\\ $(\dmatm,\dmsol)\,\times10^{-3}\text{eV}^2$\\$(\delta,\alpha)^\circ\;,\;m_{\beta\beta}\,(\text{meV})$}&$r_N$\\
		\hline
		\multirow{7}{*}{(T$_1$,R$_1$,B)}&$(21,31,12)$& $\Mnu_{22}=\Mnu_{23}$ &$r_N=\left|\dfrac{\Mnu_{22}}{\Mnu_{11}}\right|$&\multirow{3}{*}{\cmark ($3\sigma$)}&\multirow{3}{*}{\specialcell{$(34.5, 45.0, 8.41)$\\
				$(2.49, 7.56\times10^{-2})$\\
				$(269.7, 342.2)\;,\;47.8$}}&1.91\Tstrutt\Bstrutt\\ 
		\cline{2-2} \cline{4-4}\cline{7-7}
		&$(21,31,32)$&$\dfrac{\Mnu_{11}(\Mnu_{33}-\Mnu_{22})}{(\Mnu_{13})^2}=1$&$r_N=\left|\dfrac{\Mnu_{11}\Mnu_{22}}{(\Mnu_{13})^2}\right|$&&&12.00\Tstrutt\Bstrutt\\
		\cline{2-7}
		&$(21,12,32)$& $\Mnu_{11}=\Mnu_{13}$ &$r_N=\left|\dfrac{\Mnu_{22}}{\Mnu_{11}}\right|$&\multirow{3}{*}{\xmark}&\multirow{3}{*}{$-$}&$-$\Tstrutt\Bstrutt\\
		\cline{2-2} \cline{4-4}\cline{7-7}
		&$(31,12,32)$&$\dfrac{\Mnu_{22}(\Mnu_{33}-\Mnu_{11})}{(\Mnu_{23})^2}=1$&$r_N=\left|\dfrac{(\Mnu_{23})^2}{\Mnu_{11}\Mnu_{22}}\right|$&&&$-$\Tstrutt\Bstrutt\\		 
		\hline
		\multirow{7}{*}{(T$_2$,R$_1$,C)}
		&$(21,31,12)$& $\Mnu_{23}=\Mnu_{33}$, &$r_N=\left|\dfrac{\Mnu_{33}}{\Mnu_{11}}\right|$&\multirow{3}{*}{\cmark ($3\sigma$)}&\multirow{3}{*}{\specialcell{$(34.5, 45.0, 8.40)$\\
				$(2.49, 7.56\times10^{-2})$\\
				$(270.3, 17.8)\;,\;47.8$}}&1.91\Tstrutt\Bstrutt\\
		\cline{2-2} \cline{4-4}\cline{7-7}
		&$(21,31,22)$&$\dfrac{\Mnu_{11}(\Mnu_{22}-\Mnu_{33})}{(\Mnu_{12})^2}=1$&$r_N=\left|\dfrac{\Mnu_{11}\Mnu_{33}}{(\Mnu_{12})^2}\right|$&&&12.00\Tstrutt\Bstrutt\\
		\cline{2-7}
		&$(21,12,22)$& $\Mnu_{11}=\Mnu_{12}$, &$r_N=\left|\dfrac{(\Mnu_{23})^2}{\Mnu_{11}\Mnu_{33}}\right|$&\multirow{3}{*}{\xmark}&\multirow{3}{*}{$-$}&$-$\Tstrutt\Bstrutt\\
		\cline{2-2} \cline{4-4}\cline{7-7}
		&$(31,12,22)$&$\dfrac{\Mnu_{33}(\Mnu_{22}-\Mnu_{11})}{(\Mnu_{23})^2}=1$&$r_N=\left|\dfrac{\Mnu_{33}}{\Mnu_{11}}\right|$&&&$-$\Tstrutt\Bstrutt\\
		\hline
		\multirow{7}{*}{(T$_3$,R$_2$,D)}&$(11,31,12)$&\multirow{3}{*}{$\Mnu_{12}=\Mnu_{23}$}&$\dfrac{r_N-1}{r_N-\sqrt{r_N}-1}=\left|\dfrac{\Mnu_{33}}{\Mnu_{13}}\right|$&\multirow{3}{*}{\xmark}&\multirow{3}{*}{$-$}&$-$\Tstrutt\Bstrutt\\
		\cline{2-2} \cline{4-4}\cline{7-7}
		&$(11,31,22)$&&$\dfrac{\sqrt{r_N}}{r_N-1}=\left|\dfrac{\Mnu_{23}}{\Mnu_{33}}\right|$&&&$-$\Tstrutt\Bstrutt\\
		\cline{2-7}
		&$(11,12,22)$&\multirow{3}{*}{$\dfrac{\Mnu_{33}(\Mnu_{11}-2\Mnu_{12})}{(\Mnu_{13}-\Mnu_{23})^2}=1$}&$\dfrac{\sqrt{r_N}}{r_N-1}=\left|\dfrac{(\Mnu_{23})^2}{\Mnu_{12}\Mnu_{33}}\right|$&\multirow{3}{*}{\xmark}&\multirow{3}{*}{$-$}&$-$\Tstrutt\Bstrutt\\
		\cline{2-2} \cline{4-4}\cline{7-7}
		&$(31,12,22)$&&$\dfrac{\sqrt{r_N}}{r_N-1}=\left|\dfrac{\Mnu_{23}}{\Mnu_{33}}\right|$&&&$-$\Tstrutt\Bstrutt\\
		\hline
		\multirow{7}{*}{(T$_6$,R$_2$,F)}
		&$(11,21,12)$&\multirow{3}{*}{ $\Mnu_{13}=\Mnu_{23}$}&$\dfrac{r_N-1}{r_N-\sqrt{r_N}-1}=\left|\dfrac{\Mnu_{22}}{\Mnu_{12}}\right|$&\multirow{3}{*}{\xmark}&\multirow{3}{*}{$-$}&$-$\Tstrutt\Bstrutt\\		
		\cline{2-2} \cline{4-4}\cline{7-7}
		&$(11,21,32)$&&$\dfrac{\sqrt{r_N}}{r_N-1}=\left|\dfrac{\Mnu_{23}}{\Mnu_{22}}\right|$&&&$-$\Tstrutt\Bstrutt\\	
		\cline{2-7}
		&$(11,12,32)$&\multirow{3}{*}{$\dfrac{\Mnu_{22}(\Mnu_{11}-2\Mnu_{13})}{(\Mnu_{12}-\Mnu_{23})^2}=1$}&$\dfrac{\sqrt{r_N}}{r_N-1}=\left|\dfrac{(\Mnu_{23})^2}{\Mnu_{13}\Mnu_{22}}\right|$&\multirow{3}{*}{\xmark}&\multirow{3}{*}{$-$}&$-$\Tstrutt\Bstrutt\\	
		\cline{2-2} \cline{4-4}\cline{7-7}
		&$(21,12,32)$&&$\dfrac{\sqrt{r_N}}{r_N-1}=\left|\dfrac{\Mnu_{23}}{\Mnu_{22}}\right|$&&&$-$\Tstrutt\Bstrutt\\
		\hline			
		\multirow{7}{*}{(T$_1$,R$_3$,D)}&$(21,31,12)$&\multirow{3}{*}{ $\dfrac{\Mnu_{11}(\Mnu_{33}-2\Mnu_{23})}{(\Mnu_{13}-\Mnu_{12})^2}=1$}&$\dfrac{\sqrt{r_N}}{r_N-1}=\left|\dfrac{\Mnu_{12}}{\Mnu_{11}}\right|$&\multirow{3}{*}{\cmark ($3\sigma$)}&\multirow{3}{*}{\specialcell{$(37.1, 45.0, 8.46)$\\
				$(2.49, 7.56\times10^{-2})$\\
				$(347.1, 172.7)\;,\;13.2$}}&1.46\Tstrutt\Bstrutt\\
		\cline{2-2} \cline{4-4}\cline{7-7}
		&$(21,31,32)$&&$\dfrac{\sqrt{r_N}}{r_N-1}=\left|\dfrac{(\Mnu_{12})^2}{\Mnu_{11}\Mnu_{23}}\right|$&&&1.08\Tstrutt\Bstrutt\\
		\cline{2-7}
		&$(21,12,32)$&\multirow{3}{*}{$\Mnu_{12}=\Mnu_{23}$}&$\dfrac{\sqrt{r_N}}{r_N-1}=\left|\dfrac{\Mnu_{12}}{\Mnu_{11}}\right|$&\multirow{3}{*}{\xmark}&\multirow{3}{*}{$-$}&$-$\Tstrutt\Bstrutt\\
		\cline{2-2} \cline{4-4}\cline{7-7}
		&$(31,12,32)$&&$\dfrac{r_N-1}{r_N-\sqrt{r_N}-1}=\left|\dfrac{\Mnu_{11}}{\Mnu_{13}}\right|$&&&$-$\Tstrutt\Bstrutt\\
		\hline
		\multirow{7}{*}{(T$_2$,R$_3$,F)}
		&$(21,31,12)$&\multirow{3}{*}{ $\dfrac{\Mnu_{11}(\Mnu_{22}-2\Mnu_{23})}{(\Mnu_{13}-\Mnu_{12})^2}=1$}&$\dfrac{\sqrt{r_N}}{r_N-1}=\left|\dfrac{\Mnu_{13}}{\Mnu_{11}}\right|$&\multirow{3}{*}{\cmark ($3\sigma$)}&\multirow{3}{*}{\specialcell{$(36.9, 45.0, 8.46)$\\
				$(2.49, 7.56\times10^{-2})$\\
				$(188.5, 184.8)\;,\;13.2$}}&1.46\Tstrutt\Bstrutt\\		
		\cline{2-2} \cline{4-4}\cline{7-7}
		&$(21,31,22)$&&$\dfrac{\sqrt{r_N}}{r_N-1}=\left|\dfrac{(\Mnu_{13})^2}{\Mnu_{11}\Mnu_{23}}\right|$&&&1.08\Tstrutt\Bstrutt\\	
		\cline{2-7}
		&$(21,12,22)$&\multirow{3}{*}{$\Mnu_{13}=\Mnu_{23}$}&$\dfrac{r_N-1}{r_N-\sqrt{r_N}-1}=\left|\dfrac{\Mnu_{11}}{\Mnu_{12}}\right|$&\multirow{3}{*}{\xmark}&\multirow{3}{*}{$-$}&$-$\Tstrutt\Bstrutt\\
		\cline{2-2} \cline{4-4}\cline{7-7}
		&$(31,12,22)$&&$\dfrac{\sqrt{r_N}}{r_N-1}=\left|\dfrac{\Mnu_{13}}{\Mnu_{11}}\right|$&&&$-$\Tstrutt\Bstrutt\\	
		\hline\hline
	\end{tabular}
	\caption{Parameter relations (third and forth column) and low-energy predictions (sixth column) for each set of textures ($\Ynu$, $\Mr$, $\Mnu$) with three equal elements in $\Ynu$ (second column). The predicted values for the heavy neutrino mass ratio $r_N$ are also shown (last column). The results correspond to the case $\Yl=\Yl_{\rm diag}$.}
	\label{Tabrel}
\end{table*}
\begin{align}
r_N=\frac{M_2}{M_1}\,,
\label{rNdef}
\end{align}
where $M_{2,1}$ are the eigenvalues of $\Mr$, we obtain the relations shown in the fourth column of Table~\ref{Tabrel}. Our analysis shows that only eight combinations are compatible with neutrino data at the $3\sigma$ level (see fifth and sixth columns of Table~\ref{Tabrel}). The low-energy predictions for the neutrino parameters correspond to the case in which the data is best fitted. It is possible to show analytically that, for all compatible sets of matrices, $\theta_{23}=\pi/4$, which is confirmed by the numerical result. It is worth mentioning that any texture combination obtained from those presented in Table~\ref{Tabrel} by permuting the columns of $\Ynu$ will remain valid. For instance, the first case shown in Table~\ref{Tabrel} becomes (T$_4$,R$_1$,B) with equal elements (11,22,32), leading to the same predictions. Therefore, there are actually sixteen different cases compatible with the data. As mentioned above, the equality among elements of $\Ynu$ fixes the value of $r_N$, which is indicated in the last column of Table~\ref{Tabrel}. From inspection of the same table, one can also conclude that none of the texture configurations is compatible with the data at $1\sigma$.

As in the analysis presented in the previous section, the results obtained with equal $\Ynu$ elements correspond to $\Yl=\Yl_{\rm diag}$. For a different $\Yl$ texture related to $\Yl_{\rm diag}$ by permutations of lines (and columns), the $\Ynu$ textures transform among themselves, and the equal elements of $\Ynu$ change position. Thus, combinations which are incompatible with data (see Table~\ref{Tabrel}) in the charged-lepton mass basis may become compatible for a non-diagonal $\Yl$, related to $\Yl_{\rm diag}$ by permutation of lines. In Table~\ref{tabrelper} we summarize the transformation properties of each combination $({\rm T}_i,\text{R}_i,\text{A-F})$ with equal $\Ynu$ elements under line permutations $\mathbf{P}_{ij}$ (and up to possible column permutation). In each case, we identify the compatibility with data taking into account the results given in Table~\ref{Tabrel} for $\Yl=\Yl_{\rm diag}$.
\begin{table}[t]
	\centering
	\setlength\extrarowheight{2pt}
	\begin{tabular}{K{1.5cm}K{0.35cm}|K{1.5cm}K{0.35cm}|K{1.5cm}K{0.35cm}|K{1.5cm}K{0.35cm}}
		\hline\hline
		\multicolumn{2}{c}{\textbf{Texture}}&\multicolumn{2}{|c}{$\mathbf{P}_{12}$}&\multicolumn{2}{|c}{$\mathbf{P}_{13}$}&\multicolumn{2}{|c}{$\mathbf{P}_{23}$}\Tstruttt\Bstruttt\\
		\hline\hline
		\multicolumn{2}{c}{(T$_1$,R$_1$,B)}&\multicolumn{2}{|c}{(T$_1$,R$_1$,B)}&\multicolumn{2}{|c}{(T$_6$,R$_1$,E)}&\multicolumn{2}{|c}{(T$_2$,R$_1$,C)}\Tstruttt\Bstruttt\\
		\hline
		(21,31,12)&\cmark&(21,12,32)&\xmark&\multirow{4}{*}{$-$}&\multirow{4}{*}{\xmark}&(21,31,12)&\cmark\\
		(21,31,32)&\cmark&(31,12,32)&\xmark&  & &(21,31,22)&\cmark\\
		(21,12,32)&\xmark&(21,31,12)&\cmark&  & &(31,12,22)&\xmark\\
		(31,12,32)&\xmark&(21,31,32)&\cmark& & &(21,12,22)&\xmark\\
		\hline
		\multicolumn{2}{c}{(T$_2$,R$_1$,C)}&\multicolumn{2}{|c}{(T$_3$,R$_1$,E)}&\multicolumn{2}{|c}{(T$_2$,R$_1$,C)}&\multicolumn{2}{|c}{(T$_1$,R$_1$,B)}\Tstruttt\Bstruttt\\
		\hline
		(21,31,12)&\cmark&\multirow{4}{*}{$-$}&\multirow{4}{*}{\xmark}&(31,12,22)&\xmark&(21,31,12)&\cmark\\
		(21,31,22)&\cmark &&&(21,12,22)&\xmark& (21,31,32)&\cmark\\
		(21,12,22)&\xmark &&&(21,31,22)&\cmark &(31,12,32)&\xmark\\
		(31,12,22)&\xmark &&&(21,31,12)&\cmark& (21,12,32)&\xmark\\
		\hline
		\multicolumn{2}{c}{(T$_3$,R$_2$,D)}&\multicolumn{2}{|c}{(T$_2$,R$_2$,A)}&\multicolumn{2}{|c}{(T$_1$,R$_3$,D)}&\multicolumn{2}{|c}{(T$_6$,R$_2$,F)}\Tstruttt\Bstruttt\\
		\hline
		(11,31,12)&\xmark&\multirow{4}{*}{$-$}&\multirow{4}{*}{\xmark}&(31,12,32)&\xmark&(11,21,12)&\xmark\\
		(11,31,22)&\xmark& &&(21,12,32)&\xmark& (11,21,32)&\xmark\\
		(11,12,22)&\xmark& &&(21,31,32)&\cmark& (11,12,32)&\xmark\\
		(31,12,22)&\xmark &&&(21,31,12)&\cmark& (21,12,32)&\xmark\\
		\hline
		\multicolumn{2}{c}{(T$_6$,R$_2$,F)}&\multicolumn{2}{|c}{(T$_2$,R$_3$,F)}&\multicolumn{2}{|c}{(T$_1$,R$_2$,A)}&\multicolumn{2}{|c}{(T$_3$,R$_2$,D)}\Tstruttt\Bstruttt\\
		\hline
		(11,21,12)&\xmark&(21,12,22)&\xmark&\multirow{4}{*}{$-$}&\multirow{4}{*}{\xmark}&(11,31,12)&\xmark\\
		(11,21,32)&\xmark&(31,12,22)&\xmark&& &(11,31,22)&\xmark\\
		(11,12,32)&\xmark& (21,31,22)&\cmark&& &(11,12,22)&\xmark\\
		(21,12,32)&\xmark&(21,31,12)&\cmark&& &(31,12,22)&\xmark\\
		\hline
		\multicolumn{2}{c}{(T$_1$,R$_3$,D)}&\multicolumn{2}{|c}{(T$_4$,R$_3$,A)}&\multicolumn{2}{|c}{(T$_3$,R$_2$,D)}&\multicolumn{2}{|c}{(T$_2$,R$_3$,F)}\Tstruttt\Bstruttt\\
		\hline
		(21,31,12)&\cmark&\multirow{4}{*}{$-$}&\multirow{4}{*}{\xmark}&(31,12,22)&\xmark&(21,31,12)&\cmark\\
		(21,31,32)&\cmark& &&(11,12,22)&\xmark &(21,31,22)&\cmark\\
		(21,12,32)&\xmark& &&(11,31,22)&\xmark& (31,12,22)&\xmark\\
		(31,12,32)&\xmark& &&(11,31,12)&\xmark& (21,12,22)&\xmark\\
		\hline
		\multicolumn{2}{c}{(T$_2$,R$_3$,F)}&\multicolumn{2}{|c}{(T$_6$,R$_2$,F)}&\multicolumn{2}{|c}{(T$_5$,R$_3$,A)}&\multicolumn{2}{|c}{(T$_1$,R$_3$,D)}\Tstruttt\Bstruttt\\
		\hline
		(21,31,12)&\cmark&(21,12,32)&\xmark&\multirow{4}{*}{$-$}&\multirow{4}{*}{\xmark}&(21,31,12)&\cmark\\
		(21,31,22)&\cmark& (11,12,32)&\xmark&& &(21,31,32)&\cmark\\
		(21,12,22)&\xmark&(11,21,12)&\xmark&& &(31,12,32)&\xmark\\
		(31,12,22)&\xmark&(11,21,32)&\xmark&& &(21,12,32)&\xmark\\									
		\hline\hline
	\end{tabular}
	\caption{Transformation properties under the permutation matrix $\mathbf{P}_{ij}$ (permutations of the charged-lepton flavors $i$ and $j$) for the texture combination ($\Ynu$, $\Mr$, $\Mnu$) with three equal elements in $\Ynu$. The compatibility of each texture with data is also indicated considering the results shown in Table~\ref{Tabrel} for $\Yl=\Yl_{\rm diag}$. The check marks (\cmark) indicate compatibility with data at $3\sigma$.}
	\label{tabrelper}
\end{table}

\section{Leptogenesis in the 2RHNSM with texture zeros}
\label{sec3}

In the previous sections, several mass matrix patterns were found to be compatible with current neutrino oscillation data at $1\sigma$ and $3\sigma$ C.L., in the framework of the minimal type-I seesaw model with maximally restricted texture zeros. Here, we further analyze these patterns requiring their compatibility with successful leptogenesis~\cite{Fukugita:1986hr}. We recall that the baryon asymmetry of the Universe is parametrized through the baryon-to-photon ratio
\begin{gather}
\eta_B\equiv\dfrac{n_B-n_{\bar{B}}}{n_\gamma}\; ,
\label{baryontophotonratio}
\end{gather}
where $n_B$, $n_{\bar{B}}$ and $n_\gamma$ are the number densities of baryons, anti-baryons and photons, respectively. From cosmic microwave background (CMB) measurements provided by the Planck collaboration~\cite{Ade:2015xua}, the present value of $\eta_B$ is 
\begin{gather}
\eta_B^0=(6.11\pm0.04)\times 10^{-10}\,.
\label{presentetab}
\end{gather}

In a minimal type-I seesaw context with two right-handed neutrinos, the leptogenesis mechanism may proceed via the out-of-equilibrium decays of the heavy neutrinos $N_1$ and $N_2$ in the early Universe. The generated lepton asymmetry in such decays is partially converted into a baryon asymmetry by ($B+L$)-violating sphaleron processes, leading to~\cite{Antusch:2011nz}
\begin{gather}
\eta_B=a_\text{sph}\,\dfrac{N_{B-L}}{N_\gamma^\text{rec}}\simeq 9.58\times 10^{-3}\, N_{B-L}\;  ,
\label{finaletab}
\end{gather}
where $a_\text{sph} \equiv B/(B-L)=28/79$ is the conversion factor, $N_{B-L}$ is the final asymmetry calculated in a comoving volume, and $N_{\gamma}^\text{rec}$ is the number of photons in the same volume ($N_{\gamma}^\text{rec}\simeq 37.01$) at the recombination temperature. 

\subsection{Flavored and unflavored CP asymmetries}
\label{subsecCP}

An important ingredient in the generation of the BAU is the CP asymmetry produced in the decays of the heavy neutrinos into the lepton flavors $\alpha=e,\mu,\tau$. Working in the mass eigenbasis of the heavy neutrinos $N_i$ and the charged leptons $\ell_\alpha$, the CP asymmetries $\epsilon_i^\alpha$ may be computed as~\cite{Covi:1996wh}
\begin{gather}
\epsilon_i^\alpha=\dfrac{\Gamma(N_i\rightarrow\Phi\ell_\alpha)-\Gamma(N_i\rightarrow\Phi^\dagger\bar{\ell}_\alpha)}{\sum_\beta[\Gamma(N_i\rightarrow\Phi\ell_\beta)+\Gamma(N_i\rightarrow\Phi^\dagger\bar{\ell}_\beta)]},
\label{generalCP}
\end{gather}
where $\Gamma(N_i\rightarrow\Phi\ell_\alpha)\equiv \Gamma_i^\alpha$ and $\Gamma(N_i\rightarrow\Phi^\dagger\bar{\ell}_\alpha)\equiv \overline{\Gamma}_i^\alpha$ are the $N_i$ decay rates into leptons and antileptons, respectively. At tree level,
\begin{gather}
\Gamma_i^\alpha=\overline{\Gamma}_i^\alpha=M_i\dfrac{|\mathbf{Y}^\nu_{\alpha i}|^2}{16 \pi},
\label{flvdecayrate}
\end{gather}
with the sum in the denominator of~\eqref{generalCP} running over the three lepton flavors. 
The leading non-zero contributions to the asymmetry $\epsilon_i^\alpha$ arise from interference of the tree-level process with its one-loop corrections. For the two RH neutrino case, the result is~\cite{Branco:2011zb}
\begin{align}
\epsilon_i^\alpha =&\frac{1}{8\pi}\frac{1}{\mathbf{H}_{ii}^\nu}\{\text{Im}[\mathbf{Y}_{\alpha i}^{\nu *}\mathbf{H}_ {ij}^\nu\mathbf{Y}_{\alpha j}^\nu ][f(x_j)+g(x_j)]+\nonumber\\ &\text{Im}[\mathbf{Y}_{\alpha i}^{\nu *}\mathbf{H}_ {ji}^\nu\mathbf{Y}_{\alpha j}^{\nu}]g'(x_j)\},
\label{flavouredcp}
\end{align}
where $j\neq i=1,2$, $x_j=M_j^2/M_i^2$ and $\mathbf{H}^\nu=\mathbf{Y}^{\nu\dagger} \mathbf{Y}^\nu$. The loop functions $f(x)$, $g(x)$ and $g'(x)$ correspond to the one-loop vertex and self-energy corrections, given by
\begin{align}
\label{loopfunctions1}
f(x)&=\sqrt{x}\left[1-(1-x)\ln\left(1+\frac{1}{x}\right)\right],\\
g(x)&=\sqrt{x}g'(x)=-\frac{\sqrt{x}}{(x-1)}\,.
\end{align}
Summing over the lepton flavors in Eq.~\eqref{flavouredcp}, the unflavored CP asymmetry is recovered,
\begin{gather}
\epsilon_i=\frac{1}{8\pi}\frac{1}{\mathbf{H}_{ii}^\nu}\text{Im}[(\mathbf{H}_{ij}^\nu)^2][f(x_j)+g(x_j)].
\label{unflavouredcp}
\end{gather}

In our study, two temperature regimes will be of interest~\cite{Barbieri:1999ma,Abada:2006fw,Nardi:2006fx,Abada:2006ea}. For temperatures above $10^{12}$ GeV in the early Universe, the charged-lepton Yukawa interactions are out of equilibrium. Hence, for this temperature range, the three lepton flavors are indistinguishable (unflavored regime), and the lepton asymmetry may be represented rigorously by a single flavor eigenstate. In this case, the relevant CP asymmetry for leptogenesis is given by Eq.~\eqref{unflavouredcp}. In the temperature interval $10^{9}\lesssim T \lesssim 10^{12}$~GeV, the $\tau$ Yukawa interactions enter thermal equilibrium and processes involving leptons are able to distinguish between two different flavors: the $\tau$ and a coherent superposition of $e$ and $\mu$ (two-flavored regime). The corresponding CP asymmetries, $\epsilon_i^\tau$ and $\epsilon_i^\gamma\equiv \epsilon_i^e+\epsilon_i^\mu$, are then obtained from Eq.~\eqref{flavouredcp}.

The CP asymmetries given in Eq.~\eqref{flavouredcp} depend on the Yukawa coupling matrix $\Ynu$, which can be written in terms of the Casas-Ibarra parametrization presented in Eq.~\eqref{CasasandIbarra}. This allows to rewrite the asymmetry in a more convenient form for leptogenesis analysis,
\begin{widetext}
	\begin{gather}
	\epsilon_i^\alpha=-\dfrac{1}{8\pi v^2}\dfrac{M_j}{\sum_k m_k |\mathbf{R}_{ki}|^2}\sum_{k,k'}\sqrt{m_k}m_{k'}\{\sqrt{m_{k'}}\, \text{Im}[\mathbf{U}_{\alpha k}^*\mathbf{U}_{\alpha k'}\mathbf{R}_{ki}\mathbf{R}_{k'i}][f(x_j)+g(x_j)]\nonumber\\ 
	+\sum_{k''}\sqrt{m_{k''}}\,\text{Im}[\mathbf{U}_{\alpha k}^*\mathbf{U}_{\alpha k''}\mathbf{R}_{ki}\mathbf{R}_{k'i}^*\mathbf{R}_{k' j}\mathbf{R}_{k'' j}^*]g'(x_j)\}\, ,
	\label{flavouredcpCI}
	\end{gather}
\end{widetext}
where the orthogonal matrix $\mathbf{R}$ is parametrized by a single complex parameter $z$, as shown in Eq.~\eqref{RmatrixIO}. For an inverted hierarchical neutrino mass spectrum,\footnote{Hereafter, we consider only the IH case since, as shown in Section~\ref{sec2}, this is the only type of spectrum compatible with low-energy neutrino data.} the flavored asymmetries generated by $N_1$ and $N_2$ decays are written in terms of $z$ as
\begin{align}
\label{flavouredCPtanz1}
\epsilon_1^{\alpha}=&-\dfrac{M_2}{8\pi v^2}\dfrac{ A_1^\alpha\left[f(x_2) + g(x_2)\right]+B_1^\alpha g'(x_2)}{m_1|c_z|^2+m_2|s_z|^2}\, ,\\
\epsilon_2^{\alpha}=&-\dfrac{M_1}{8\pi v^2}\dfrac{ A_2^\alpha\left[f(x_1) + g(x_1)\right]+B_2^\alpha g'(x_1)}{m_1|s_z|^2+m_2|c_z|^2}\, ,
\label{flavouredCPtanz2}
\end{align}
where $c_z\equiv\cos z$, $s_z\equiv \sin z$ and
\begin{align}
\label{Afactor}
A_1^\alpha=&(m_2^2|\mathbf{U}_{\alpha 2}|^2-m_1^2|\mathbf{U}_{\alpha 1}|^2)\, \text{Im}[s^2_z]+\xi\sqrt{m_1 m_2}\nonumber\\
&\lbrace(m_2-m_1)\text{Im}[\textbf{U}_{\alpha 1}^*\textbf{U}_{\alpha 2}]\text{Re}[c_z s_z]+\nonumber\\
&+(m_2+m_1)\text{Re}[\mathbf{U}^*_{\alpha 1}\mathbf{U}_{\alpha 2}]\text{Im}[c_z s_z]\rbrace\, ,\\
\nonumber\\
B_1^\alpha=&m_1m_2\,(|\mathbf{U}_{\alpha 2}|^2-|\mathbf{U}_{\alpha 1}|^2)\, \text{Im}[c^2_z\, (s^2_z)^*]+\xi\sqrt{m_1 m_2}\nonumber\\
&\lbrace(\,|c_z|^2+|s_z|^2\,)(m_2-m_1)\text{Im}[\textbf{U}_{\alpha 1}^*\textbf{U}_{\alpha 2}]\text{Re}[c_z\, s_z^*]+\nonumber\\
&+(\,|c_z|^2-|s_z|^2\,)(m_2+m_1)\text{Re}[\mathbf{U}^*_{\alpha 1}\mathbf{U}_{\alpha 2}]\text{Im}[c_z\, s_z^*]\rbrace.\nonumber\\
\label{Bfactor}
\end{align}
The factors $A_2^\alpha$ and $B_2^\alpha$ are obtained replacing $s_z\rightarrow c_z$, $c_z\rightarrow s_z$ and $\xi\rightarrow-\xi$ in Eqs.~\eqref{Afactor} and \eqref{Bfactor}, respectively. These factors have the following properties
\begin{gather}
\sum_{\alpha}A_1^\alpha=\dmsol\text{Im}[s_z^2], \quad \sum_{\alpha}A_2^\alpha=\dmsol\text{Im}[c_z^2]\, ,\nonumber\\
\sum_{\alpha}B_i^\alpha=0\, .
\end{gather}
Using these relations, the unflavored CP asymmetries~\eqref{unflavouredcp} are easily obtained,
\begin{gather}
\label{unflavouredcpCIz1}
\epsilon_1=-\dfrac{M_2}{8\pi v^2}\dfrac{\dmsol\text{Im}[s_z^2]}{m_1\,|c_z|^2 + m_2\, |s_z|^2}\left[f(x_2)+g(x_2)\right],\\
\epsilon_2=-\dfrac{M_1}{8\pi v^2}\dfrac{\dmsol\text{Im}[c_z^2]}{m_1\,|s_z|^2 + m_2\, |c_z|^2}\left[f(x_1)+g(x_1)\right].
\label{unflavouredcpCIz2}
\end{gather}

The presence of a texture zero in $\Ynu$ allows for the determination of $z$ in terms of low-energy parameters and $M_{1,2}$, as one may see from Eq.~\eqref{CasasandIbarra}. For instance, in the basis where the charged-lepton and RH neutrino mass matrices are diagonal, the condition $\Ynu_{11}=0$ implies, for IH,
\begin{gather}
\sqrt{m_1}\,\U_{11}^*c_z+\xi\sqrt{m_2}\,\U_{12}^*s_z=0\; ,
\end{gather}
leading to
\begin{gather}
\tan z=-\xi\sqrt{\dfrac{m_1}{m_2}}\dfrac{\U_{11}^*}{\U_{12}^*}\; .
\end{gather}
In Table~\ref{tabletan}, we present the expressions for $\tan z$ according to the position of the texture zero in $\Ynu$ and considering the matrix forms R$_{1,2,3}$ for $\Mr$. From this table it is straightforward to see that requiring the presence of two simultaneous zeros in $\Ynu$ leads to relations among the mixing angles, neutrino masses and the low-energy phases, as expected from Eq.~\eqref{IHrelation}.
\begin{table*}[t!]
	\centering
	\setlength\extrarowheight{2pt}
	\begin{tabular}{K{1cm}|K{5cm}|K{5cm}}
		\hline\hline
		$\Mr$&$\tan z$ for $\Ynu_{\alpha 1}=0$&$\tan z$ for $\Ynu_{\alpha 2}=0$\\
		\hline
		R$_1$&$-\xi\sqrt{\dfrac{m_1}{m_2}}\dfrac{\mathbf{U}_{\alpha 1}^*}{\mathbf{U}_{\alpha 2}^*}$&$\xi\sqrt{\dfrac{m_2}{m_1}}\dfrac{\mathbf{U}_{\alpha 2}^*}{\mathbf{U}_{\alpha 1}^*}$\Tstrut\Bstrut\\
		R$_2$&$i$&$\dfrac{- i\, \sqrt{m_1}M_1\mathbf{U}^*_{\alpha 1}+\,\xi\sqrt{m_2}M_2\mathbf{U}^*_{\alpha 2}}{\sqrt{m_1}M_2\mathbf{U}^*_{\alpha 1}+\,i\,\xi\sqrt{m_2}M_1\mathbf{U}^*_{\alpha 2}}$\Tstrut\Bstrut\\
		R$_3$&$\dfrac{i\, \sqrt{m_1}M_1\mathbf{U}^*_{\alpha 1}+\,\xi\sqrt{m_2}M_2\mathbf{U}^*_{\alpha 2}}{\sqrt{m_1}M_2\mathbf{U}^*_{\alpha 1}-\,i\,\xi\sqrt{m_2}M_1\mathbf{U}^*_{\alpha 2}}$& $i$\Tstrut\Bstrut\\
		\hline\hline
	\end{tabular}
	\caption{Expressions for $\tan z$ as a function of the low-energy parameters and the heavy-neutrino masses $M_1$ and $M_2$, for each texture in the IH case.}
	\label{tabletan}
\end{table*}
Replacing in Eqs.~\eqref{flavouredCPtanz1} and \eqref{flavouredCPtanz2} the expressions for $\tan z$ given in Table~\ref{tabletan},  and using the low-energy relations of Table~\ref{CPtable}, we obtain predictions for the flavored CP asymmetries $\epsilon_i^\tau$ and $\epsilon_i^\gamma$, for each of the valid texture-zero cases identified in the Section~\ref{sec2}. 
\begin{figure}[t!]
	\hspace*{-0.2cm}
	\includegraphics[scale=0.8]{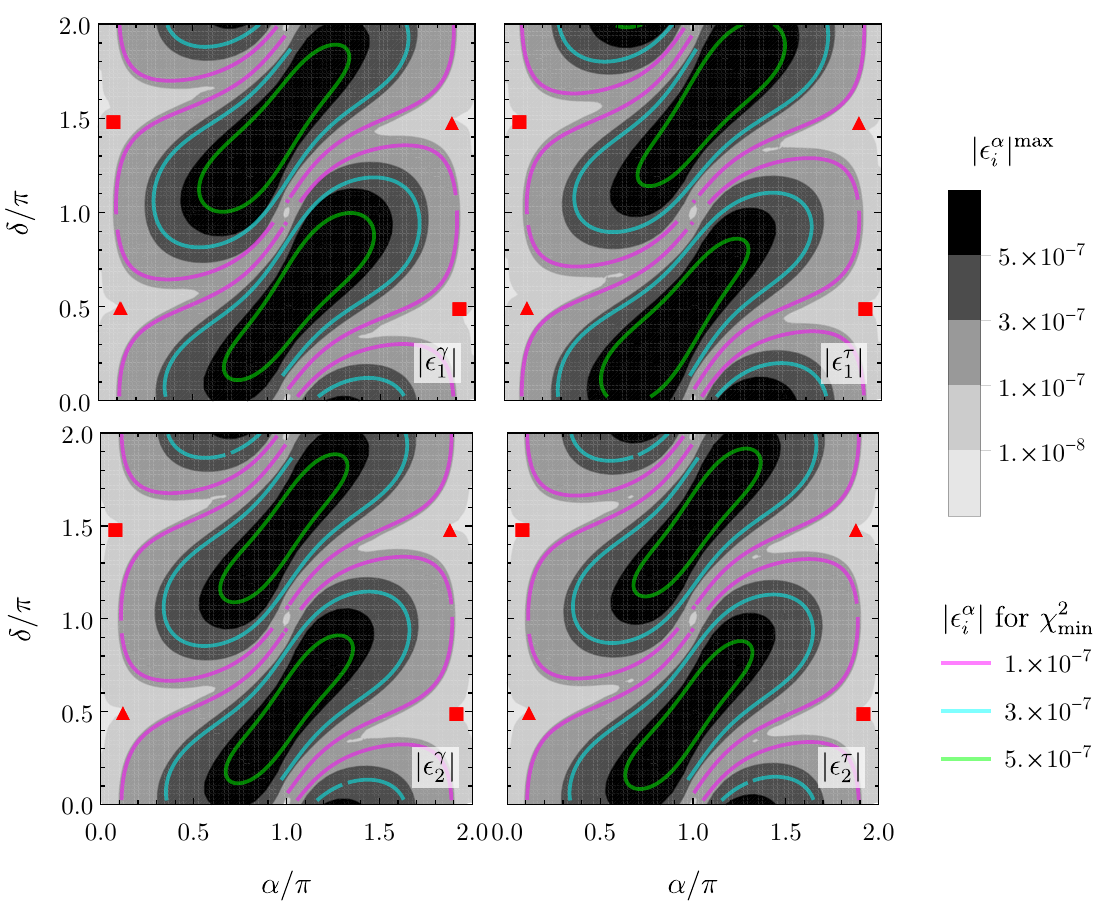} \\
	\caption{Flavored CP asymmetries $|\epsilon_{1,2}^\gamma|$ and $|\epsilon_{1,2}^\tau|$ as functions of the low-energy CP-violating phases $\alpha$ and $\delta$, for the texture-zero case $\Ynu_{11}=0$ and R$_1$. The gray-scale contour regions show the maximum values of $|\epsilon_i^\alpha|$, taking $\theta_{ij}$, $\dmsol$ and $|\dmatm|$ in the $3\sigma$ experimental range (see Table~\ref{datatable}) and for $10^9\lesssim M_{1,2}\lesssim 10^{12}$~GeV with $M_2\gtrsim 3M_1$. The colored contour lines are the results obtained for the minimum value of $\chi^2$. In the plot, the triangles and squares correspond to the ($\alpha$,$\delta$) pairs fixed by the conditions $\Ynu_{11}=\Ynu_{22}=0$ and $\Ynu_{11}=\Ynu_{32}=0$, respectively (cf. textures B and C in Table~\ref{tabR1R2R3}).}
	\label{cpasymflav}
\end{figure}

It turns out that, even if one considers a single texture zero in $\Ynu$, the CP asymmetries are highly suppressed in the flavored regime. As illustration, in Fig.~\ref{cpasymflav} we show the asymmetries $|\epsilon_i^\gamma|$ and $|\epsilon_i^\tau|$, $i=1,2$, for the case R$_1$ and $\Ynu_{11}=0$ on the plane ($\alpha$,$\delta$) of the low-energy CP-violating phases. The maximum value for the CP asymmetries (gray scale) is presented for the $3\sigma$ range of the mixing angles and the neutrino mass-squared differences. Notice that we have imposed $M_2\gtrsim 3M_1$ to ensure a nonresonant regime, and $10^9\lesssim M_{1,2}\lesssim 10^{12}$~GeV since $\mu$ and $e$ interactions are in equilibrium. In the same plot, the $|\epsilon_i^\alpha|$ values calculated for the minimum of $\chi^2$ (varying the mixing angles and mass-squared differences) are presented as colored lines. The points marked by triangles and squares correspond to ($\alpha$,$\delta$) fixed by the two-zero conditions $\Ynu_{11}=\Ynu_{22}=0$ and $\Ynu_{11}=\Ynu_{32}=0$, respectively, i.e. textures B and C for $\Mnu$ (see Table~\ref{tabR1R2R3}). We may also see that for the whole $\delta$ and $\alpha$ ranges, the obtained CP asymmetries are highly suppressed being the maximum values below $10^{-6}$. Moreover, $|\epsilon_i^\alpha|\lesssim 10^{-7}$ for ($\alpha$,$\delta$) fixed by textures B and C. Thus, for the case with $\Ynu_{11}=0$ and $R_1$, the CP asymmetries are too small to ensure efficient leptogenesis. One can show that all other combinations of textures with zeros in $\Ynu$ and $\Mr$ allowed by neutrino data yield similar results. 

We conclude that thermal leptogenesis in the flavored regime with $10^9\lesssim T\lesssim 10^{12}$~GeV cannot successfully reproduce the observed baryon asymmetry given in Eq.~\eqref{presentetab}. This conclusion will be corroborated in the next section when the final baryon asymmetry is computed. 

\begin{figure}[t!]
	\includegraphics[scale=0.92]{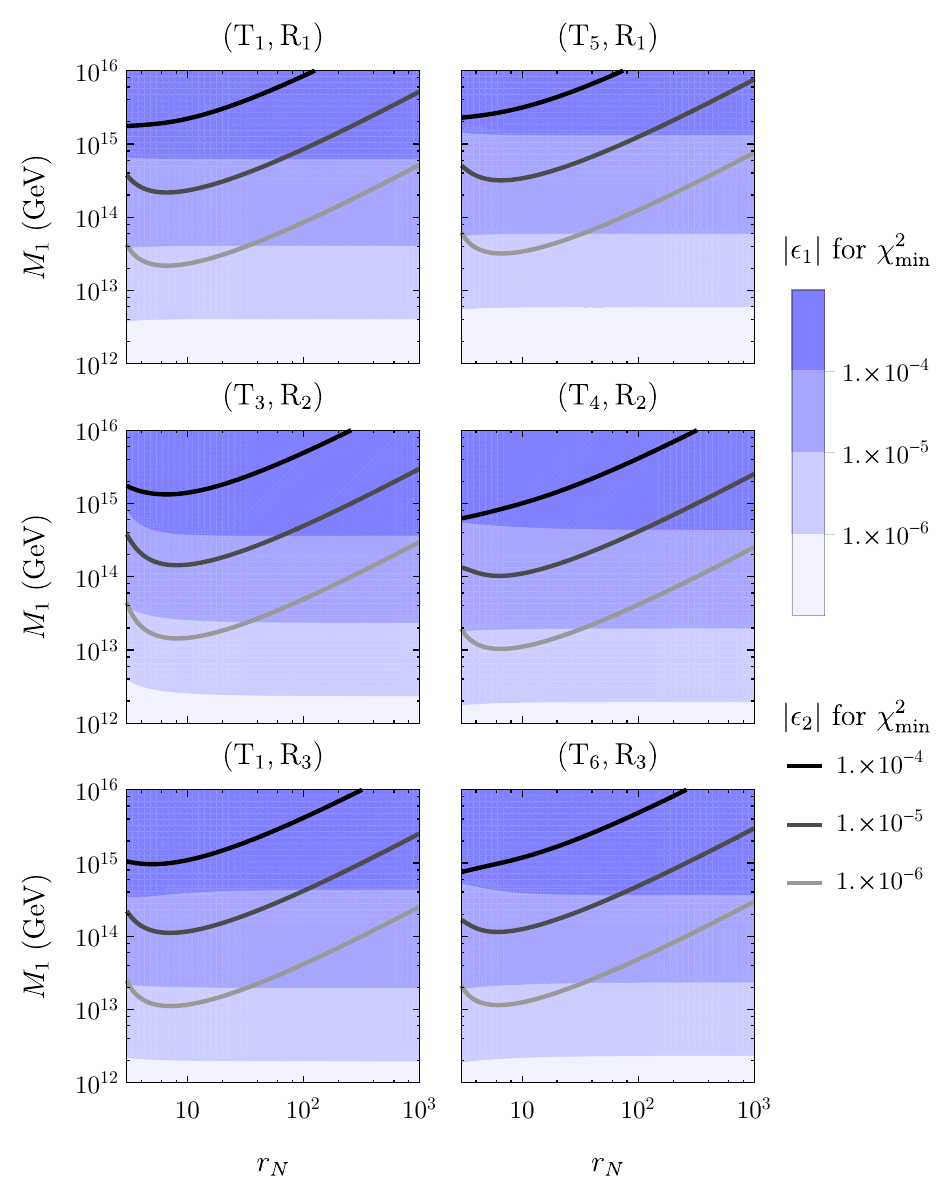} \\
	\caption{Unflavored CP asymmetries $|\epsilon_i|$, $i=1,2$ on the plane ($r_N$, $M_1$), $r_N= M_2/M_1$, for the low-energy neutrino parameters that best fit the texture pairs (T,R). The blue contour regions (gray-scale contour lines) show $|\epsilon_1|$ ($|\epsilon_2|$).}
	\label{cpasymunflav}
\end{figure}

Let us consider now the unflavored regime. In this case, the CP asymmetries~\eqref{unflavouredcp} are enhanced. For each of the valid two-zero textures, the  CP asymmetries $\epsilon_1$ and $\epsilon_2$ given in Eqs.~\eqref{unflavouredcpCIz1} and \eqref{unflavouredcpCIz2} are computed using the expressions of Tables~\ref{tabletan} and \ref{CPtable}. In Fig.~\ref{cpasymunflav}, we present $|\epsilon_1|$ (blue contour regions) and $|\epsilon_2|$ (gray-scale contour lines) in the $(r_N,M_1)$ plane, for the low-energy neutrino parameters that best fit the 2RHNSM with $\Ynu$ and $\Mr$ textures (T,R). We only show the results for the six combinations (T$_{1,5}$,R$_1$), (T$_{3,4}$,R$_2$), and (T$_{1,6}$,R$_3$), that lead to $\eta_B>0$. From the same plot we see that the maximum values for $|\epsilon_i|$ can now reach $10^{-4}$, which is two orders of magnitude higher than the ones in the flavored regime (cf. Fig~\ref{cpasymflav}).  Furthermore, as the ratio $r_N$ increases, the CP asymmetry $|\epsilon_2|$ gets slightly suppressed with respect to $|\epsilon_1|$. 

\subsection{Baryon asymmetry production}


In the calculation of the final lepton asymmetry we will consider the contributions of both $N_1$ and $N_2$. In the flavored and unflavored regimes, the leptonic CP asymmetries generated in the $N_i$ decays are most likely to be washed out by the out-of-equilibrium inverse decays and scattering processes in which the heavy neutrinos participate. In general, a measure of the washout strength is given by the so-called decay parameter $K_i$, which for a lepton flavor channel $\alpha$ reads
\begin{gather}
K_i^\alpha=\dfrac{\tilde{m}_i^\alpha}{m_*},
\label{flvdecayparam}
\end{gather}
where $\tilde{m}_i^\alpha$ is the flavored effective neutrino mass,
\begin{gather}
\tilde{m}_i^\alpha=\dfrac{
	v^2|\Ynu_{\alpha i}|^2}{M_i},
\end{gather}
and $m_*\simeq 1.09\times 10^{-3}$~eV is the equilibrium neutrino mass. Summing over flavors in Eq.~\eqref{flvdecayparam}, one obtains the total decay parameter, 
\begin{gather}
K_i=\sum_\alpha K_i^\alpha=\frac{\tilde{m}_i}{m_*},
\end{gather}
with
\begin{gather}
\tilde{m}_i=\sum_\alpha \tilde{m}_i^\alpha=\frac{v^2\,\mathbf{H}^\nu_{ii}}{M_i}.
\end{gather}
The relation between $\tilde{m}_i$ and $m_*$ gives a measure of thermal equilibrium for the decays, namely, if $\tilde{m}_i\gg m_*$  ($\tilde{m}_i\ll m_*$) the asymmetry is strongly (weakly)  washed out by inverse decays. 

The fraction of surviving lepton asymmetry can be expressed in terms of efficiency factors $\kappa\in[0,1]$, which are obtained by solving the relevant Boltzmann equations. In our study, we will use instead the simple and accurate analytical approximations for $\kappa_i^\alpha(K_i^\alpha)$ and $\kappa_i(K_i)$ from  Refs.~\cite{Antusch:2011nz} and \cite{Buchmuller:2004nz}, respectively. The imposed hierarchy $M_2\gtrsim 3 M_1$ implies $N_{N_1}(T\sim M_2)\simeq N_{N_2}(T\sim M_1)\simeq 0$, so that the computation of the final asymmetry may be split into the $N_1$ and $N_2$ leptogenesis phases. Furthermore, we consider a strong-coupling $N_1$ scenario, where part of the lepton asymmetry generated by $N_2$ decays is projected onto a flavor-direction protected against the washout from $N_1$ interactions~\cite{Antusch:2011nz}.

The final ($B-L$)-asymmetry for the flavored temperature regime can be written as~\cite{Antusch:2011nz}
\begin{gather}
N_{B-L}=N_{\Delta_{\gamma_1}}+N_{\Delta_{\gamma_1^\perp}}+N_{\Delta_{\tau}},
\label{finalasym}
\end{gather}
where the $\Delta_\alpha\equiv B/3-L_\alpha$ number densities in each flavor state read
\begin{align}
N_{\Delta_{\gamma_1\;}}&\simeq -P_{\gamma_2\gamma_1}\,\epsilon_2^\gamma\,\kappa_2^\gamma\, e^{-\frac{3\pi}{8}K_1^\gamma}-\epsilon_1^\gamma\,\kappa_1^\gamma,\\
N_{\Delta_{\tau\;\;\;}}&\simeq -\epsilon_2^\tau\,\kappa_2^\tau\, e^{-\frac{3\pi}{8}K_1^\tau}-\epsilon_1^\tau\,\kappa_1^\tau,\\
N_{\Delta_{\gamma_1^\perp}}&\simeq -\,(1-P_{\gamma_2\gamma_1})\,\epsilon_2^\gamma\,\kappa_2^\gamma,
\end{align}
in which $\gamma_1$ and $\gamma_1^\perp$ are the parallel and orthogonal flavor components to the interaction channels of $N_1$, respectively. Here, $\kappa_i^\alpha$ are the efficiency factors defined in~\cite{Antusch:2011nz}, and $P_{\gamma_2\gamma_1}$ is the probability of flavor $\gamma_2$, generated in the $N_2$ decay, to be transformed into $\gamma_1$ under the $N_1$ decay process,
\begin{align}
P_{\gamma_2\gamma_1}=\dfrac{\left|\sum_{\alpha} \mathbf{Y}^{\nu *}_{\alpha 1} \Ynu_{\alpha 2}\right|^2}{\left(\,\sum_{\alpha} |\Ynu_{\alpha 1}|^2\right) \left(\, \sum_{\alpha} |\Ynu_{\alpha 2}|^2 \right)},
\end{align}
where $\alpha=e, \mu$.

In the unflavored regime, the lepton flavors are indistinguishable in the primordial plasma and the final $(B-L)$-asymmetry reads~\cite{Blanchet:2011xq}
\begin{gather}
\label{unflavouredbaryonasym}
N_{B-L}\simeq-\epsilon_1\kappa_1-\left(1-P_{21}+P_{21}e^{-\frac{3\pi K_1}{8}}\right)\epsilon_2\kappa_2\; ,
\end{gather}
with $\kappa_i$ being defined in~\cite{Buchmuller:2004nz}. Here, $P_{21}$ is the probability of the lepton asymmetry produced in $N_2$ leptogenesis being projected onto the flavor direction of the asymmetry due to $N_1$ interactions,
\begin{align}
P_{21}=\dfrac{\left|\mathbf{H}^\nu_{12}\right|^2}{\mathbf{H}^\nu_{11} \mathbf{H}^\nu_{22}}.
\end{align}
After computing the densities $N_{B-L}$, for both flavored and unflavored regimes, using Eqs.~\eqref{finalasym} and \eqref{unflavouredbaryonasym}, the final baryon-to-photon ratio $\eta_B$ is obtained from Eq.~\eqref{finaletab}.
\begin{figure}[t!]
	\hspace*{-0.2cm}
	\includegraphics[scale=0.8]{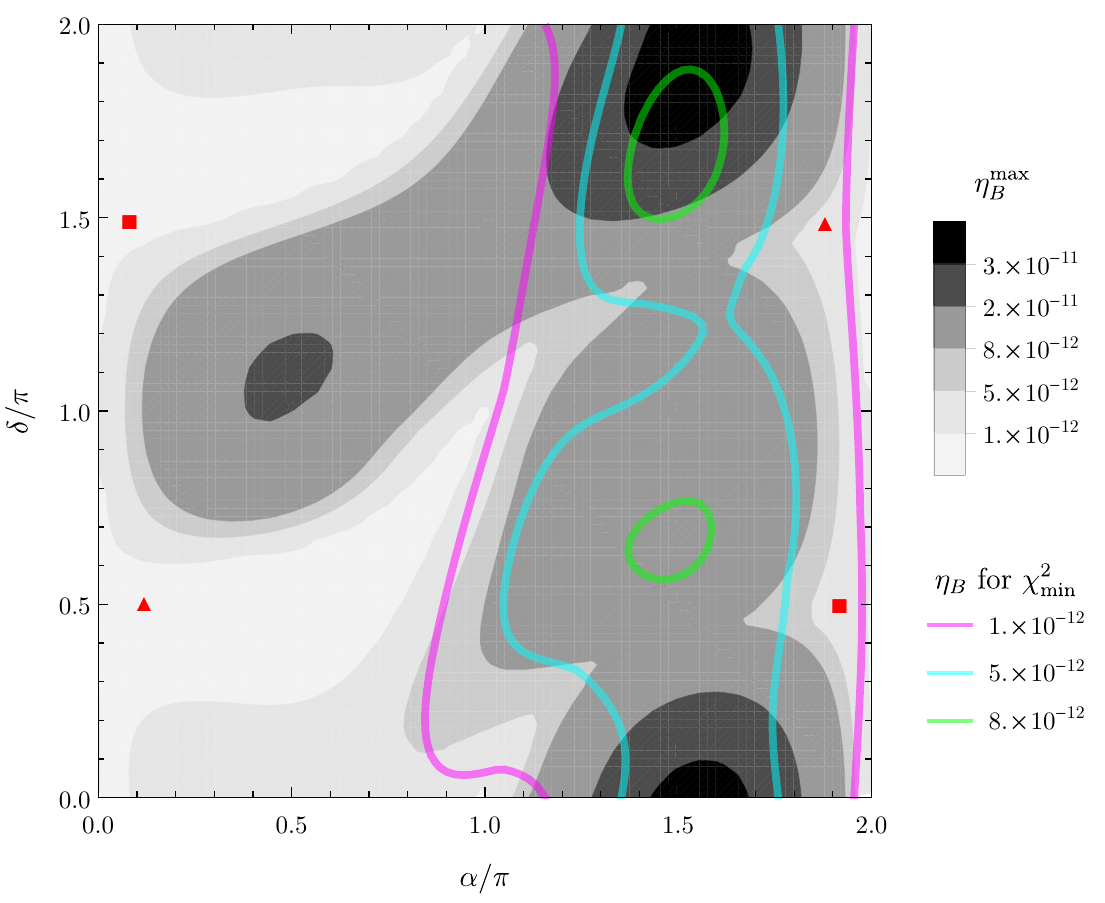} \\
	\caption{Baryon-to-photon ratio $\eta_B$ as a function of the low-energy CP-violating phases $\alpha$ and $\delta$ in the flavored regime, for the texture-zero case $\Ynu_{11}=0$ and R$_1$. The gray-scale contour regions show the maximum value of $\eta_B$, taking $\theta_{ij}$, $\dmsol$ and $|\dmatm|$ in the $3\sigma$ experimental range (see Table~\ref{datatable}) and for $10^9\lesssim M_{1,2}\lesssim 10^{12}$~GeV with $M_2\gtrsim 3M_1$. The colored contour lines are the results obtained for the minimum value of $\chi^2$. In the plot, the triangles and squares correspond to the ($\alpha$,$\delta$) pairs fixed by the conditions $\Ynu_{11}=\Ynu_{22}=0$ and $\Ynu_{11}=\Ynu_{32}=0$, respectively (cf. textures B and C in Table~\ref{tabR1R2R3}).}
	\label{etabflav}
\end{figure}

In Fig.~\ref{etabflav}, we present $\eta_B$ computed for the illustrative case of $\Ynu_{11}=0$ with R$_1$, for which the flavored CP asymmetries were already analyzed in Section~\ref{subsecCP}. In that figure, the gray-scale contour regions correspond to the maximum of $\eta_B$ in the $3\sigma$ experimental range of the mixing angles and the neutrino mass-squared differences, taking $10^9\lesssim M_{1,2}\lesssim10^{12}$~GeV. As expected from the small values of $|\epsilon_i^\alpha|$ (see Fig.~\ref{cpasymflav}), the final baryon asymmetry is suppressed in the whole allowed parameter region. Indeed, the final $\eta_B$ lies between one to two orders of magnitude below the observed value $\eta_B^0$. Moreover, for the $\Mnu$ textures B and C, marked in the figure by a triangle and a square, respectively, $\eta_B\lesssim10^{-12}$ is verified. For all the other combinations of textures T and R that are compatible with neutrino oscillation data, similar results were obtained for the flavored regime, corroborating the fact that thermal leptogenesis in the two-flavor case is not viable. 

\begin{figure}[t!]
	\includegraphics[scale=0.92]{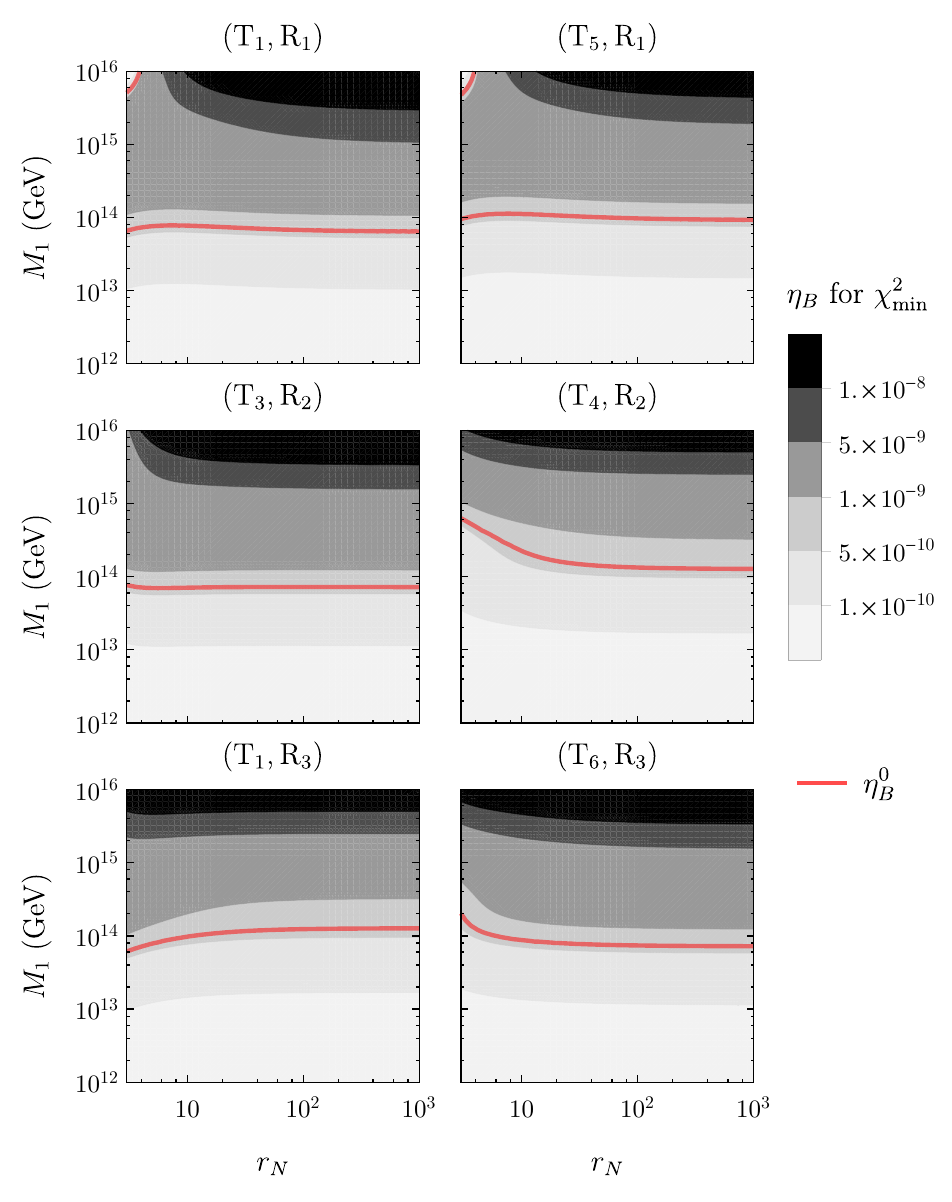} \\
	\caption{The baryon-to-photon ratio $\eta_B$ on the plane ($r_N$, $M_1$), $r_N= M_2/M_1$, for the unflavored regime and taking the low-energy neutrino parameters that best fit the texture pairs (T,R). The gray-scale contour regions represent the final value of $\eta_B$, while the red contour line corresponds to the observed value $\eta_B^0$ given in Eq.~\eqref{presentetab}.}
	\label{etabunflav}
\end{figure}

For the unflavored regime, sufficiently large (and positive) values for $\eta_B$ are obtained for six of the twelve pairs (T,R) of textures compatible with neutrino data (see Table~\ref{tabR1R2R3}). This is shown in Fig.~\ref{etabunflav}, where we present the predicted $\eta_B$ (gray-scale contour regions) as a function of $M_1$ and the mass ratio $r_N$, considering the low-energy neutrino data that best fit the six textures. In fact, for all these cases, the observed baryon-to-photon ratio $\eta_B^0$ (red contour line in Fig.~\ref{etabunflav}) is achieved for $M_1\sim 10^{14}$~GeV, being $\kappa_i\sim \mathcal{O}(10^{-3})$ (strong washout regime). Hence, one concludes that the texture combinations (T$_{1,5}$,R$_1$), (T$_{3,4}$,R$_2$), and (T$_{1,6}$,R$_3$) lead to successful thermal leptogenesis in the unflavored regime.

\begin{figure*}[t!]
	\includegraphics[scale=0.75]{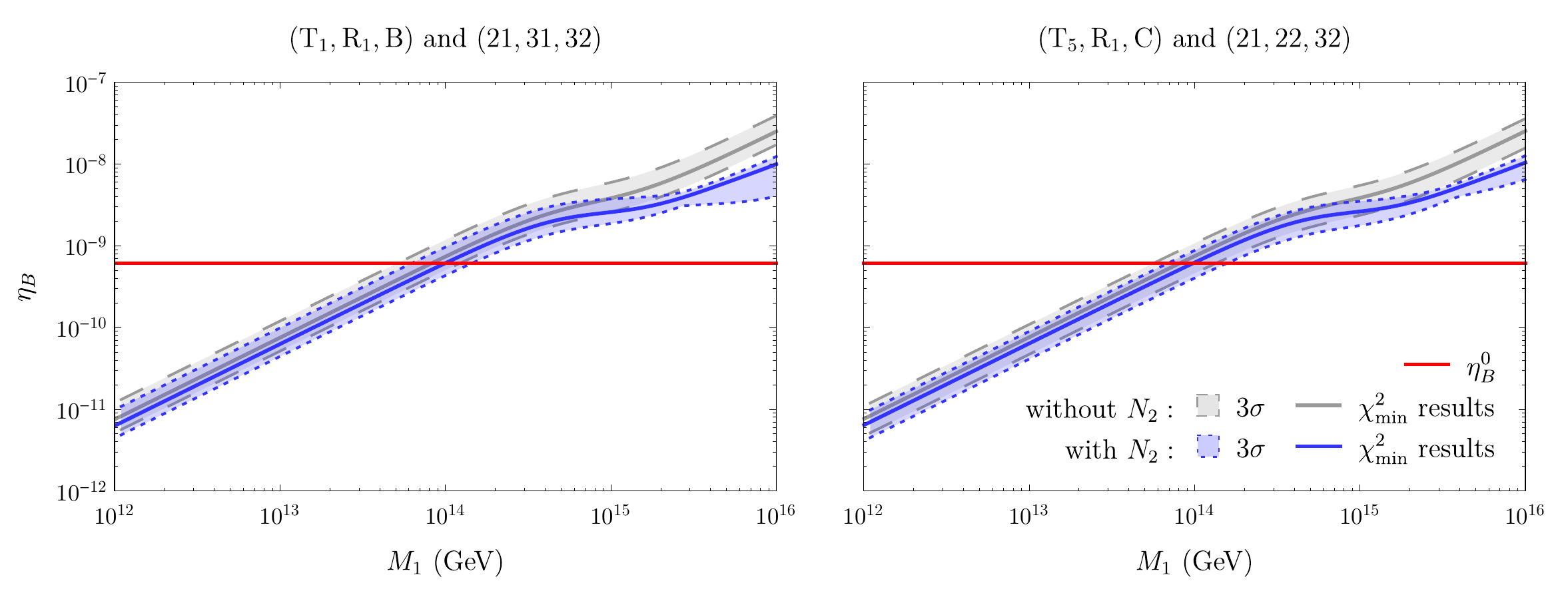} \\
	\caption{Baryon-to-photon ratio $\eta_B$ as a function of $M_1$ for the cases (T$_1$,R$_1$,B) with $\Ynu_{21}=\Ynu_{31}=\Ynu_{32}$, on the left, and (T$_5$,R$_1$,C) with $\Ynu_{21}=\Ynu_{22}=\Ynu_{32}$, on the right, using the 3$\sigma$ range for $\theta_{ij}$, $\dmsol$ and $|\dmatm|$ given in Table~\ref{datatable}. The blue (gray) region corresponds to the case where the contribution of $N_2$ to the final asymmetry is (not) accounted for. The solid blue and gray lines are the $\eta_B$ predictions obtained using the low-energy parameters that best fit the considered textures (see Table~\ref{Tabrel}). The horizontal red line represents the present baryon-to-photon ratio $\eta_B^0$.}
	\label{etabcorr}
\end{figure*}

One may wonder whether the above conclusion remains valid, if one considers the more restricted cases discussed in Section~\ref{sec3a}, in which three elements of $\Ynu$ are equal. We will only consider the cases that were proved to be compatible with neutrino data and, additionally, verify the condition $r_N\gtrsim3$, for which our leptogenesis assumptions hold. From Table~\ref{Tabrel} and Fig.~\ref{etabunflav}, one can see that only the cases (T$_1$,R$_1$,B) with $\Ynu_{21}=\Ynu_{31}=\Ynu_{32}$ and (T$_5$,R$_1$,C) with $\Ynu_{21}=\Ynu_{22}=\Ynu_{32}$ meet those requirements ($r_N\sim 12$) and, simultaneously, yield $\eta_B>0$. In Fig.~\ref{etabcorr}, we present the $\eta_B$ region allowed by the $3\sigma$ experimental interval for the low-energy neutrino parameters (blue region) as a function of the mass $M_1$. Here we also show the results obtained when the contribution of the second neutrino $N_2$ is not taken into account for leptogenesis (gray region). One concludes that for temperatures below $10^{14}$~GeV the effect of the second neutrino $N_2$ is negligible, while for higher temperatures the $N_2$ contribution tends to lower $\eta_B$. The value of $\eta_B^0$ (red horizontal line) is achieved for masses $M_1\sim 10^{14}$~GeV. 

\section{Conclusions}
\label{sec4}

In this paper, we have revisited the 2RHNSM considering maximally restricted texture-zero patterns for the lepton Yukawa and mass matrices. Our results are summarized in Table~\ref{tabR1R2R3}. We conclude that textures B, C and D for the effective neutrino mass matrix $\Mnu$ are compatible with current neutrino data (mixing angles and mass-squared differences) at $1\sigma$, while texture F is compatible at $3\sigma$. In all cases, only an inverted hierarchical neutrino mass spectrum is allowed. A remarkable prediction of textures B and C is that one of the viable solutions for the low-energy CP-violating Dirac phase is $\delta\sim 3\pi/2$, which is very close to the best-fit value obtained from the combined fit of neutrino oscillation data. 

Aiming at reducing the number of free parameters in the model, we have also explored scenarios in which additional relations (equality) among the Dirac neutrino Yukawa couplings are imposed. The cases with the maximum number of equal elements in $\Ynu$ which are compatible with neutrino data are presented in Table~\ref{Tabrel}. As can be seen from the table, compatibility is only verified at the $3\sigma$ confidence level.

For the phenomenologically viable textures, we have studied their implications for the BAU in the framework of type-I seesaw thermal leptogenesis. We paid special attention to the treatment of leptogenesis in the 2RHNSM. Contrary to what is customary in the literature, where only the decay of the lightest heavy neutrino is considered, we included the decays of both heavy neutrinos in our analysis. Moreover, flavor effects that arise from the fact that lepton interactions exit thermal equilibrium at different temperatures in the early Universe were taken into account. We considered two temperature regimes for leptogenesis: the two-flavored regime ($10^9\lesssim T\lesssim 10^{12}$~GeV) and the unflavored regime ($T\gtrsim10^{12}$~GeV). Within our assumptions ($M_2\gtrsim 3M_1$), we showed that the CP asymmetries in the flavored regime are too small to generate the required lepton asymmetry for successful leptogenesis. On the other hand, for the unflavored case, the CP asymmetries are enhanced, and the observed baryon-to-photon ratio is achieved in the 2RHNSM for the texture combinations (T$_{1,5}$,R$_1$), (T$_{3,4}$,R$_2$), and (T$_{1,6}$,R$_3$), for $M_1\sim 10^{14}$~GeV. Furthermore, the cases (T$_1$,R$_1$) and (T$_5$,R$_1$), with three equal elements in $\Ynu$ in the positions (21,31,32) and (21,22,32), respectively, were shown to be also compatible with the present value of the baryon asymmetry for the same leptogenesis temperature $T \sim 10^{14}$~GeV.

The nature of the flavor structure of the fermion sector in the standard model and theories beyond it remains puzzling. A common approach to address this problem is to assume certain constraints on the coupling and/or mass matrices in order to reduce the number of free parameters. The lepton textures considered in this work were taken as the simplest and most economical patterns that can be implemented in the framework of the 2RHNSM. We have shown that the maximally constrained 2RHNSM is compatible with current neutrino oscillation data and can also explain the matter-antimatter asymmetry in the Universe via the leptogenesis mechanism. This conclusion holds for several mass matrix textures with the maximal number of allowed zeros and, in a more restricted set, having equal elements in the Dirac Yukawa coupling matrix. It would be interesting to see if such predictive textures could arise from a flavor symmetry principle. This is a subject that certainly deserves to be further explored~\cite{inprep}.
\\

{\bf Acknowledgements:} This work was supported by Funda\c{c}{\~a}o para a Ci{\^e}ncia e a Tecnologia (FCT, Portugal) through the project CFTP-FCT Unit 777 (UID/FIS/00777/2013).

\end{document}